\documentclass[prb,aps,twocolumn,10pt,superscriptaddress,notitlepage,longbibliography,nofootinbib]{revtex4-1}
\usepackage{graphicx}
\usepackage{amsmath,mathtools}
\usepackage{bbold}
\usepackage{amssymb}
\usepackage{epstopdf}
\usepackage{siunitx}

\usepackage[dvipsnames]{xcolor}

\usepackage{textgreek}
\usepackage[ngerman,english]{babel}
\usepackage{comment}

\usepackage[caption=false]{subfig}
\usepackage{bm}

\usepackage{hyperref}
 \hypersetup{
     colorlinks=true,
     linkcolor=blue,
     filecolor=blue,
     citecolor = magenta,      
     urlcolor=red,
     }

\usepackage{braket}
\usepackage{xcolor}

\usepackage{xargs}
\newcommandx{\greencom}[2][1=]
{\todo[inline, color=green!40,#1]{#2}}
\newcommandx{\bluecom}[2][1=]
{\todo[inline, color=blue!40,#1]{#2}}
\newcommandx{\bluemargin}[2][1=]
{\todo[color=blue!40,#1]{#2}}

\usepackage{letltxmacro}
\LetLtxMacro{\ORIGselectlanguage}{\selectlanguage}
\makeatletter
\DeclareRobustCommand{\selectlanguage}[1]{%
  \@ifundefined{alias@\string#1}
    {\ORIGselectlanguage{#1}}
    {\begingroup\edef\x{\endgroup
       \noexpand\ORIGselectlanguage{\@nameuse{alias@#1}}}\x}%
}
\newcommand{\definelanguagealias}[2]{%
  \@namedef{alias@#1}{#2}%
}

\makeatother

\definelanguagealias{en}{english}
\definelanguagealias{EN}{english}
\definelanguagealias{eng}{english}
\definelanguagealias{de}{ngerman}

\begin{document}

\title{Fluctuation-dissipation theorem and  fundamental photon commutation relations in lossy nanostructures using quasinormal modes}

\author{Sebastian Franke}
\email{sebastian.franke@tu-berlin.de}
\affiliation{Technische Universit\"at Berlin, Institut f\"ur Theoretische Physik,
Nichtlineare Optik und Quantenelektronik, Hardenbergstra{\ss}e 36, 10623 Berlin, Germany}
 \author{Juanjuan Ren}
 \affiliation{\hspace{0pt}Department of Physics, Engineering Physics, and Astronomy, Queen's University, Kingston, Ontario K7L 3N6, Canada\hspace{0pt}}
 \author{Stephen Hughes}
\affiliation{\hspace{0pt}Department of Physics, Engineering Physics, and Astronomy, Queen's University, Kingston, Ontario K7L 3N6, Canada\hspace{0pt}}
  
\author{Marten Richter}
\affiliation{Technische Universit\"at Berlin, Institut f\"ur Theoretische Physik,
 Nichtlineare Optik und Quantenelektronik, Hardenbergstra{\ss}e 36, 10623 Berlin, Germany}

\date{\today}

\begin{abstract}

We provide theory and formal insight on the Green function quantization method for absorptive and dispersive spatial-inhomogeneous media in the context of dielectric media.
We show that a fundamental Green function identity, which appears, e.g., in the fundamental commutation relation 
of the electromagnetic fields, is also valid in the limit of non-absorbing media.
We also demonstrate how
the zero-point field fluctuations yields a non-vanishing surface term in configurations without absorption, when using a more formal procedure of the Green function quantization method.  
We then apply the presented method to a recently developed theory of photon quantization using quasinormal modes [Franke {\it et al.}, 
 Phys. Rev. Lett. {\bf 122}, 213901 (2019)] for finite nanostructures embedded in a lossless background medium. We discuss the strict dielectric limit of the commutation relations of the quasinormal mode operators and present different methods to obtain them, connected to the radiative loss for non-absorptive but open resonators. We show exemplary calculations of a fully three-dimensional photonic crystal beam cavity,
 including the lossless limit, which
 supports a single quasinormal mode and discuss the 
 limits of the commutation relation for vanishing damping (no material loss and no radiative loss).

\end{abstract}

\maketitle

\section{Introduction\label{Sec: Intro}}

Cavity-QED phenomena in photonics nanostructures and nanolasers, such as metal nanoparticles~\cite{nanogold2,Akselrod2016,david,NanoMarten, theuerholz2013influence} or semiconductor and dielectric microvavities~\cite{micropillars,micropillars2,Reithmaier_Nature_432_197_2004,cao2015dielectric,PhysRevB.75.115331}, 
have become an important and rising field in the research area of quantum optics and quantum plasmonics over the last few decades, since it provides a suitable platform to study, e.g., non-classical light effects~\cite{faraon2008coherent,brooks2012non} and quantum information processes~\cite{quantinfo,loss1998quantum}.
A rigorous quantum optics theory for these systems is of great importance to describe the underlying mechanisms and applications of light-matter interaction in these dissipative systems. Over two decades ago, a seminal phenomenological quantization approach for general absorbing and dispersive spatial-inhomogeneous media~\cite{Dung} based on earlier work of Huttner and Barnett~\cite{HuttnerBarnett, barnett1992spontaneous} as well as Hopfield~\cite{hopfield1958theory} was introduced. The theory has already been successfully applied to many technologically interesting quantum optical scenarios, e.g., input-output in multilayered absorbing structures~\cite{gruner1996quantum}, active quantum emitters in the vicinity of a metal sphere~\cite{scheel1999spontaneous, dung2000spontaneous},  the vacuum Casimir effect~\cite{philbin2011casimir},
strong coupling effects in
quantum plasmonics~\cite{VanVlack2012,Hmmer2013},
and non-Markovian dynamics
in nonreciprocal environments~\cite{PhysRevA.99.042508}.

A few years after the introduction of the Green function quantization scheme, the method was further confirmed by approaches using a canonical quantization formalism in combination with a Fano-type diagonalization~\cite{suttorp2004field, philbin2010canonical}. While it was 
confirmed that, for the case of absorptive bulk media, the theory is consistent with the fundamental axioms of quantum mechanics, e.g., the preservation of the fundamental commutation relations between the electromagnetic field operators~\cite{Dung, grunwel}, it has
been debated recently, whether the theory can be applied to non-absorbing media or finite-size absorptive media, e.g., dielectric nanostructures in a non-absorbing background medium~\cite{dorier2019canonical, dorier2019critical}. This is an important limit, in which results of the well-known quantization of lossless dielectrics as in the seminal approach from Ref.~\onlinecite{motu2} should be recovered. One of the reasons for 
an apparent problem to recover this limit is that the electric field operator in the formulation in Refs.~\onlinecite{Dung, grunwel} is proportional to the imaginary part of the dielectric permittivity which, at first sight, seems to vanish in the case of non-absorbing media, 
and would be inconsistent with the limiting case of quantization in free space. While it was shown explicitly for one-dimensional systems, that the Green function quantization is indeed consistent with the case of non-absorbing media~\cite{grunwel}, there were only some arguments for the general case, based on the fact that one has to include a small background absorption until the very end of the calculations~\cite{philbin2011casimir,drezet2017equivalence}.

Recently, a fully three-dimensional quantization scheme for quasinormal modes (QNMs) was presented on the basis of the mentioned Green function quantization approach and successfully applied to metal resonators and metal-dielectric hybrid resonators coupled to a quantum emitter and embedded in a lossless background medium~\cite{PhysRevLett.122.213901}. The QNMs are solutions to the Helmholtz equation with open boundary conditions and have complex eigenfrequencies, describing loss as an inherent mode property. The classical QNM approach has been used successfully to describe light-matter interaction on the semi-classical level~\cite{muljarovPert,KristensenHughes,SauvanNorm,NormKristHughes,Lalanne_review,carlson2019dissipative},
and the introduction of a mode expansion in the Green function quantization approach has the benefit to describe general dissipative systems with few modes instead of the continua used in the seminal Green function quantization approach.
At the fundamental level, the quantization of such modes is essential to study and
understand the physics of few quanta light sources~\cite{Hughes_SPS_2019,
fernandez-dominguez_plasmon-enhanced_2018}
and aspects of quantum fluctuations
of quantum nonlinear optics.
The implicit condition, that one leaves a small imaginary part of the permittivity in the equations until the end of the calculations, was also used in Ref.~\onlinecite{PhysRevLett.122.213901} and led to a contribution that describes the radiative loss of the QNMs and is finite in the case of pure dielectric structures. In this way, the final results can equally be applied to lossy
or lossless resonators that can be described by few QNMs, or indeed
a combination of both.

In this work, we 
shed fresh insights on a few of these apparent problems, and show explicitly that
there are no problems with taking the lossless limit of the Green function quantization as long as the limits are performed carefully.
Specifically
we derive, using the limit of vanishing absorption, 
an alternative formalism of the Green function quantization approach,
which helps to clarify the underlying physics and 
can be used to clearly see how to quantize open-cavity modes with and without material losses.  The theory can thus be consistently used for a wide range of problems in quantum optics and quantum plasmonics, including finite nanostructures in a lossless background medium. To show the importance of our modified approach, we apply the procedure to a rigorous QNM quantization scheme for finite nanostructures. Using this quantized QNM approach, we will show, explicitly, why the commutation relations between different QNM operators are connected to two dissipation processes in general; nonradiative loss into the absorptive medium (material loss) and radiative loss into the far field (radiation loss), which renders the theory also suitable for dielectric lossy resonators. In the limit of no material loss,
we obtain a physically meaningful result in terms of the normalized power flow
from the QNM fields. We also recover well known results for a lossless
infinite medium.

The rest of our paper is organized as follows: In Section~\ref{Sec: Theory}, we first introduce the ``system'' and construct a sequence of geometry configurations, that fully satisfies the outgoing boundary conditions of any open cavity system. With this model, we reformulate the quantization scheme from Refs.~\onlinecite{grunwel, Dung, Scheel, scheel1998qed, vogel2006}. 
In Section~\ref{Sec: FDTheorem}, we then rederive
a Green function identity, connected to the fundamental commutation relations of the electromagnetic fields, and explicitly show the appearance of a surface term in the zero-point vacuum fluctuations 
for the medium-assisted electromagnetic field operators in the case of the constructed permittivity sequences.
 In  Section~\ref{Sec: FormQuantQNM}, we recapitulate the QNM approach and adopt the modified Green function quantization approach to the quantization scheme for QNMs from Ref.~\onlinecite{PhysRevLett.122.213901}.
In Section~\ref{Sec: Commutation}, we derive the fundamental commutation relations of the QNM operators connected to the radiative dissipation. First, we apply the method from Section~\ref{Sec: FDTheorem} directly as a special case to the QNM commutation relation.
Second, we derive the far field limit of the commutation relation on the basis of two different methods (the Dyson approach and the field equivalence principle), to obtain the fields outside. 
 Finally, we presents a practical example of a three-dimensional resonator, supporting a single QNM on a
finite-loss and lossless photonic crystal beam. We discuss the numerically obtained QNM commutation relation in the limit of vanishing absorption, and show how loss impacts the various quantization factors.
We then discuss the key equations in Section~\ref{Sec: keyequations} and summarize the main results of the work in Section~\ref{Sec: Conc}. The main part of our paper is followed by five appendices, showing two more detailed derivations connected to a fundamental Green function identity, the derivation of the limit of the QNM commutation relation for vanishing dissipation, a discussion on the commutation relation of the electric and magnetic field operator using the QNM expansion, and details on the numerical calculations of the photonic crystal beam cavity.

\section{Quantization approach for lossy and absorptive media using permittivity sequences\label{Sec: Theory}}

\subsection{System and introduction of permittivity sequences\label{SubSec: Sequence}}
\begin{figure}[h]
 \centering
 \includegraphics[width=0.8\columnwidth,angle=0]{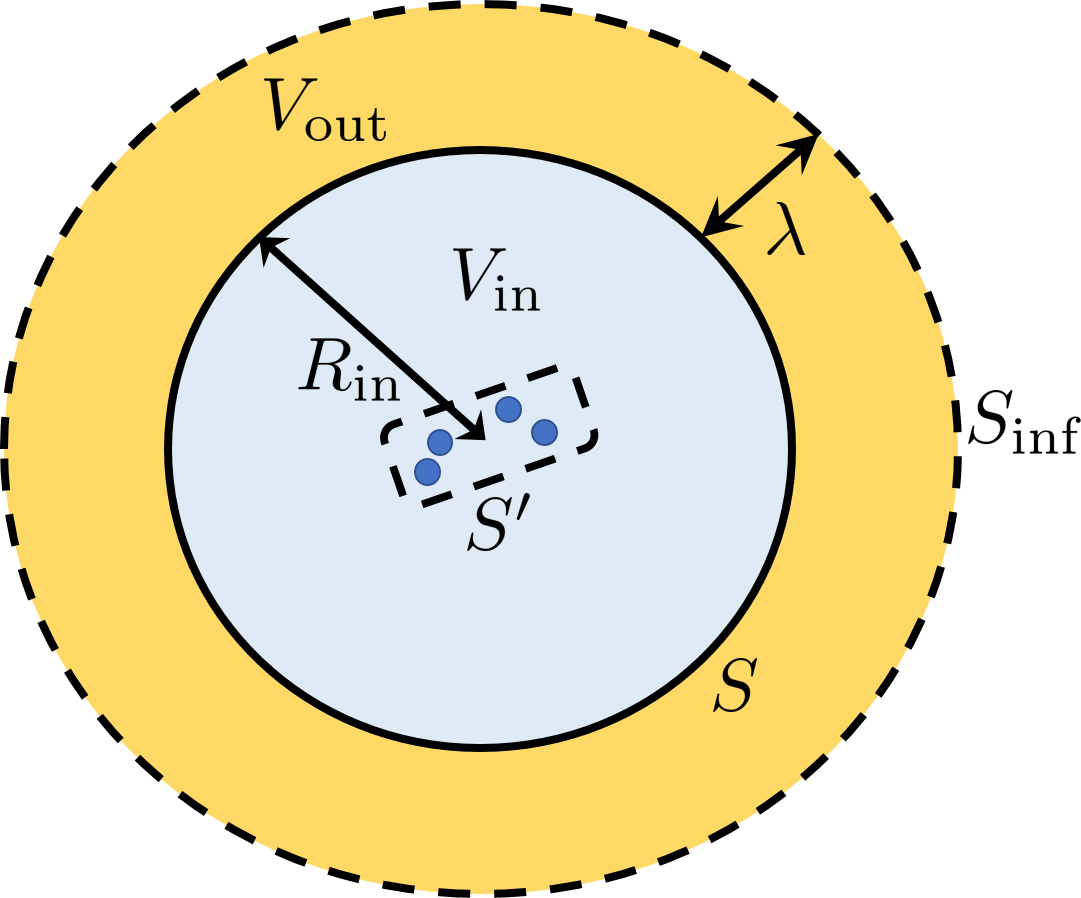}
 \vspace{-0.2cm}
 \caption{{
 Geometry of the complete system of interest. The scattering sources are contained in a spherical volume $V_{\rm in}$ with radius $R_{\rm in}$ and boundary $\mathcal{S}$. 
 $V_{\rm in}$ is surrounded by a spherical shell $V_{\rm out}(\lambda)$ of thickness $\lambda$, filled with a spatial-homogeneous medium and bounded by the surfaces $\mathcal{S}$ and $\mathcal{S}_{\rm inf}(\lambda)$.}}
\label{fig1}
\end{figure}

To keep our model as general and realistic as possible, we investigate a spatial-inhomogeneous geometry described by the Kramers-Kronig permittivity $\epsilon(\mathbf{r},\omega)=1+\chi_{\rm s}(\mathbf{r},\omega)$ (or complex dielectric constant), which can be separated into two main regions as depicted in Fig.~\ref{fig1}. The volume $V_{\rm in}$ is a spherical volume with radius $R_{\rm in}$, which shall include all scattering sources (e.g., metallic or dielectric nano resonators), described by the susceptibility $\chi_{\rm s}(\mathbf{r},\omega)$, and is otherwise filled with vacuum, described by the spatial-homogeneous background permittivity $\epsilon_{\rm B} = 1$. Furthermore, $V_{\rm in}$ is bounded by an artificial fixed spherical surface $\mathcal{S}$. The volume $V_{\rm out}(\lambda)$ is a spherical shell with variable thickness $\lambda$ filled with spatial-homogeneous background medium $\epsilon_{\rm B}=1$, 
surrounding $V_{\rm in}$ and is bounded by $\mathcal{S}$ and $\mathcal{S}_{\infty}(\lambda)$. 
We also define $V(\lambda)=V_{\rm in}+V_{\rm out}(\lambda)$  and take $V \rightarrow \mathbb{R}^3$ (all 3D space, and without any artificial surrounding surface in the far field) for the limit $\lambda\rightarrow \infty$. 

We introduce $\epsilon^{(\alpha)} (\mathbf{r},\omega)$ (describing the geometry) as the sequence of permittivity functions. Quite generally, one could add any artificial spatial-homogeneous Kramers-Kronig susceptibility $\chi_\alpha(\omega)$ with $\lim_{\alpha\rightarrow 0}\chi_\alpha(\omega)=0$ to $\epsilon(\mathbf{r},\omega)$ to construct such a sequence. However, for convenience,
we add an artificial Lorentz oscillator function to the permittivity $\epsilon(\mathbf{r},\omega)$, such that the sequence of permittivity functions $\epsilon^{(\alpha)}(\mathbf{r},\omega)$ converge for $\alpha\rightarrow 0$ to the actual permittivity, so that
\begin{equation}
\epsilon^{(\alpha)}(\mathbf{r},\omega)=\epsilon(\mathbf{r},\omega) + \alpha \chi(\omega),
\end{equation}
where $\chi(\omega)$ is the susceptibility 
\begin{equation}
\chi(\omega) = \frac{\chi_0}{\omega_0^2 - \omega^2 - i\Gamma\omega}\label{eq: SmallLorentz}, 
\end{equation}
which describes a single Lorentz oscillator with width $\Gamma$, with center frequency $\omega_0$ far off-resonant to the relevant frequencies \footnote{Note, in general this is not a requirement as we will take $\alpha \rightarrow 0$.},
and $\alpha\geq 0$. We note, that $\epsilon^{(\alpha)}(\mathbf{r},\omega)$ fulfills the Kramers-Kronig relations for all $\alpha$, 
since both, $\epsilon(\mathbf{r},\omega)$ and $\chi(\omega)$ fulfill the respective Kramers-Kronig relations. In particular, the sequence of spatial-homogeneous background permittivity functions reads
\begin{equation}
\epsilon^{(\alpha)}_{\rm B}(\omega)=1+\alpha\frac{\chi_0}{\omega_0^2 - \omega^2 - i\Gamma\omega}.
\end{equation}

\subsection{Green function quantization approach with permittivity sequences\label{SubSec: GFquantMod}}
In Refs.~\onlinecite{grunwel, Dung, scheel1998qed}, the inclusion of an artificial noise term in the Maxwell equations was proposed to preserve the fundamental spatial commutation relation between the electromagnetic field operators in the case of absorbing background media. The associated inhomogeneous Helmholtz equation for the medium-assisted electric field operator with the modified permittivity model from above then reads
\begin{equation}
\boldsymbol{\nabla}\times\boldsymbol{\nabla}\times\hat{\mathbf{E}}_{\alpha}(\mathbf{r},\omega)-\frac{\omega^2}{c^2}\epsilon^{(\alpha)}(\mathbf{r},\omega)\hat{\mathbf{E}}_{\alpha}(\mathbf{r},\omega){=} i\omega\mu_0 \hat{\mathbf{j}}_{\rm N,\alpha}(\mathbf{r},\omega),\label{eq: HelmholtzEfield}
\end{equation}
where $\hat{\mathbf{j}}_{\rm N,\alpha}(\mathbf{r},\omega)$ is a phenomenological introduced noise current density of the form 
\begin{equation}
\hat{\mathbf{j}}_{\rm N,\alpha}(\mathbf{r},\omega) = \omega\sqrt{\frac{\hbar\epsilon_0}{\pi}\epsilon_I^{(\alpha)}(\mathbf{r},\omega)}\hat{\mathbf{b}}_\alpha(\mathbf{r},\omega),\label{eq: jnoise_alpha}
\end{equation}
and $\hat{\mathbf{b}}_\alpha(\mathbf{r},\omega),\hat{\mathbf{b}}_\alpha^\dagger(\mathbf{r},\omega)$ are annihilation and creation operator of the medium-assisted electromagnetic field. Equation~\eqref{eq: HelmholtzEfield} also has the 
formal solutions 
\begin{align}
\hat{\mathbf{E}}_{\alpha}(\mathbf{r},\omega) =& \hat{\mathbf{E}}_{\alpha}^{\rm hom}(\mathbf{r},\omega)\nonumber\\
&+\frac{i}{\epsilon_0\omega}\lim_{\lambda\rightarrow \infty}\int_{V(\lambda)}{\rm d}\mathbf{r}'\mathbf{G}_{\alpha}(\mathbf{r},\mathbf{r}',\omega)\cdot \hat{\mathbf{j}}_{\rm N,\alpha}(\mathbf{r}',\omega),\label{eq: FormSolE}
\end{align}
where $\hat{\mathbf{E}}_{\alpha}^{\rm hom}(\mathbf{r},\omega)$ is the solution to the homogeneous Helmholtz equation and the second term is the scattering solution. Here, $\mathbf{G}_{\alpha}(\mathbf{r},\mathbf{r}',\omega)$ is the 
relevant Green function, fulfilling
\begin{align}
\boldsymbol{\nabla}_{\mathbf{r}}\times\boldsymbol{\nabla}_{\mathbf{r}}\times\mathbf{G}_{\alpha}(\mathbf{r},\mathbf{r}',\omega)-&\frac{\omega^2}{c^2}\epsilon^{(\alpha)}(\mathbf{r},\omega)\mathbf{G}_{\alpha}(\mathbf{r},\mathbf{r}',\omega) \nonumber\\
&= \frac{\omega^2}{c^2}\mathbb{1}\delta(\mathbf{r}-\mathbf{r}').\label{eq: GF_Helmholtz}
\end{align}
As discussed in Ref.~\onlinecite{scheel1998qed}, in order to preserve the fundamental commutation relation between the electromagnetic field operators, the only allowed solution of the homogeneous problem is the trivial zero solution. This is true for the sequence of homogeneous solutions, i.e., $\hat{\mathbf{E}}_{\alpha}^{\rm hom}(\mathbf{r},\omega)=0$.

Now, in the limit of $\alpha\rightarrow 0$, the quantities associated to the original permittivity describing the physical problem of interest are preserved. 

It is important to note that different orders of the two limits, namely ($\alpha\rightarrow 0, \lambda\rightarrow \infty$), can yield a different results; 
also note the direct application of $\alpha\rightarrow 0$ (under the spatial integral) leads to a vanishing electric field operator in the limit of non-absorbing media in $V({\lambda})$ (note the integral kernel is proportional to $(\epsilon_I^{(\alpha)}(\mathbf{r},\omega))^{1/2}$). However, 
as we will show in the next section, the prior application of the limit of unbounded volume $\lambda\rightarrow \infty$ leads to a physical contribution associated with the vacuum fluctuations on a finite boundary, that survives also in the limit of non-absorbing media. The first (direct) application of $\alpha\rightarrow 0$ corresponds to a point-wise convergence~\cite{reed2012methods}, while the first application of $\lambda \rightarrow \infty$ corresponds to a convergence  in the sense of tempered distributions. Therefore, the exchange of the limits is non-trivial and we will later clarify how to correctly carry out these limits, and how to obtain physically meaningful results.

Note that with respect to the phenomenological quantization approach, other models have also 
adopted the approach of  leaving a small imaginary part of the permittivity until the very end of the calculations~\cite{scheel1998qed, drezet2017equivalence};
below we will present a more detailed 
 mathematical treatment, and one which can easily be applied
 to quantized mode theories (using QNMs) in a rigorous and intuitive way.

\section{Green function identity and fluctation-dissipation relation for finite scattering objects\label{Sec: FDTheorem}
}

Important physical quantities in the context of macroscopic QED are the so-called zero-point field fluctuations, described by $\langle0|\hat{\mathbf{E}}(\mathbf{r}',\omega)\hat{\mathbf{E}}^\dagger(\mathbf{r},\omega)|0\rangle$, where $|0\rangle$ is the vacuum state, implicitly defined within the Green function framework via $\hat{\mathbf{b}}_\alpha(\mathbf{r},\omega)|0\rangle = 0$.  
In fact, for $\mathbf{r}=\mathbf{r}'$, it is well known that the zero-point field fluctuations are directly connected to the spontaneous emission rate of a quantum emitter at position $\mathbf{r}$, which interacts with its electromagnetic medium~\cite{dung2000spontaneous,scheel1999spontaneous}. In the following, we will first show the consistency of the modified approach with the formulas obtained in Refs.~\onlinecite{Dung,dung2000spontaneous,scheel1999spontaneous} 
and then calculate a form for finite scattering objects, which gives two contributions. The first contribution is a volume integral covering the scattering regions (which vanishes in the limit of no absorption), and
the second contribution is a surface integral (which remains finite also in non-absorptive cases). 

Using the formal solution of the sequence of electric field operators from Eq.~\eqref{eq: FormSolE} and the the bosonic nature of $\hat{\mathbf{b}}_\alpha(\mathbf{r},\omega),\hat{\mathbf{b}}_\alpha^\dagger(\mathbf{r},\omega)$, we obtain
\begin{align}
\frac{\pi\epsilon_0}{\hbar}&\int_0^\infty{\rm d}\omega'\langle0|\hat{\mathbf{E}}_\alpha(\mathbf{r},\omega)\hat{\mathbf{E}}^\dagger_\alpha(\mathbf{r}', \omega')|0\rangle\nonumber\\
&=\lim_{\lambda\rightarrow \infty}\int_{V(\lambda)}{\rm d}\mathbf{s}~ \epsilon_I^{(\alpha)}(\mathbf{s},\omega)\mathbf{G}_{\alpha}(\mathbf{r}, \mathbf{s},\omega) \cdot \mathbf{G}_{\alpha}^*(\mathbf{s}, \mathbf{r}',\omega)\nonumber\\
&\equiv \mathbf{M}_{\alpha}(\mathbf{r},\mathbf{r}',\omega)  .\label{eq: FDTRelation}
\end{align}

Equation~\eqref{eq: FDTRelation} relates the fluctuation of the electric field (LHS) to the dissipation in terms of the absorption $\epsilon_I$.
Note also that this expression is related to other important
quantities in Green function quantization, 
e.g., the fundamental commutation relation $[\hat{\mathbf{E}}_\alpha(\mathbf{r}),\hat{\mathbf{B}}_\alpha(\mathbf{r}')]$ between the electric field and the magnetic field operator~\cite{Scheel}. We will show that, in the limit $\alpha\rightarrow 0$, the Green function identity 
\begin{equation}
    \lim_{\alpha\rightarrow 0}\mathbf{M}_{\alpha}(\mathbf{r},\mathbf{r}',\omega)={\rm Im}\left[\mathbf{G}(\mathbf{r},\mathbf{r}',\omega)\right],
\end{equation}
is obtained, where $\mathbf{G}(\mathbf{r},\mathbf{r}',\omega)$ solves the Helmholtz equation, Eq.~\eqref{eq: GF_Helmholtz}, with permittivity $\epsilon(\mathbf{r},\omega)$.

In a first step, we note that $\epsilon_I=(\epsilon-\epsilon^*)/(2i)$ and use the Helmholtz equation Eq.~\eqref{eq: GF_Helmholtz} in combination with the dyadic-dyadic Green second identity (see App.~\ref{App: ProofMalpha}), to get
\begin{align}
\mathbf{M}_{\alpha}&(\mathbf{r},\mathbf{r}')={\rm Im}\left[\mathbf{G}_{\alpha}(\mathbf{r},\mathbf{r}')\right] \nonumber \\
&+\frac{c^2}{2i\omega^2}\lim_{\lambda\rightarrow\infty}\int_{\mathcal{S}_{\infty}(\lambda)}{\rm d}A_\mathbf{s}\left\{\mathbf{C}_{\alpha}(\mathbf{s},\mathbf{r},\mathbf{r}'){-} \mathbf{C}_{\alpha}^\dagger(\mathbf{s},\mathbf{r}',\mathbf{r})\right\},\label{eq: GFrel2prime}
\end{align}
with
\begin{equation}
\mathbf{C}_{\alpha}(\mathbf{s},\mathbf{r},\mathbf{r}')=\left[\mathbf{n}_\mathbf{s}\times\mathbf{G}_{\alpha}(\mathbf{s},\mathbf{r})\right]^{\rm T} \cdot \left[\boldsymbol{\nabla}_{\mathbf{s}}\times\mathbf{G}_{\alpha}^*(\mathbf{s}, \mathbf{r}')\right]\label{eq: C_Def},
\end{equation}
and 
we have dropped the explicit $\omega$ notation as an argument of the functions in the following to simplify the notation. Here, $\mathbf{n}_\mathbf{s}$ is normal vector on the surface $\mathcal{S}_{\infty}(\lambda)$, pointing outwards of $V(\lambda)$.
We focus on the integral over the spherical surface $\mathcal{S}_\infty(\lambda)$, and change to spherical coordinates with respect to $\mathbf{s}$.  Using the Dyson equation in combination with analytical properties of the Green function in spatial-homogeneous media (cf. App.~\ref{App: GFHom}), we arrive at 
\begin{align}
&\lim_{\lambda\rightarrow\infty}\int_{\mathcal{S}_{\infty}(\lambda)}{\rm d}A_\mathbf{s}~ \left\{\mathbf{C}_{\alpha}(\mathbf{s},\mathbf{r},\mathbf{r}')- \mathbf{C}_{\alpha}^\dagger(\mathbf{s},\mathbf{r}',\mathbf{r})\right\} \nonumber\\
&=\lim_{\lambda\rightarrow\infty}e^{-2k_{\alpha,I}(R_{\rm in}+\lambda)}\mathbf{I}_{\rm g,\alpha}(\mathbf{r},\mathbf{r}'),\label{eq: SinfInt}
\end{align}
with $k_{\alpha,I}=(k_{\alpha}-k_{\alpha}^*)/(2i)$, and the radius $R_{\rm in}$ of $V_{\rm in}$ and the $\mathbf{I}_{\rm g,\alpha}(\mathbf{r},\mathbf{r}')$ is a geo\-metric $\lambda$-independent factor and implicitly defined via Eq.~\eqref{eq: GfarDys}, \eqref{eq: RotGfarDys}, Eq.~\eqref{eq: Lfunctions} and Eq.~\eqref{eq: C_Def}. We further note, that $k_{\alpha}^2=\omega^2\epsilon_{\rm B}^{(\alpha)}(\omega)/c^2$.

Performing the limit 
$\lambda\rightarrow\infty$ on Eq.~\eqref{eq: SinfInt}, we see that the integral on the LHS Eq.~\eqref{eq: SinfInt} vanishes for any $\alpha>0$, since $k_{\alpha,I}>0$ for $\omega,\alpha>0$ by construction of $\chi(\omega)$ (cf. Eq.~\eqref{eq: SmallLorentz}). 
Also note that we have used here the positive root of the complex number $\epsilon_{\rm B}^{(\alpha)}(\omega)$.

The procedure and ordering of limits is connected to the weak convergence of tempered distributions~\cite{reed2012methods}. To analyze this in more detail,  
in the case of finite scattering objects with lossless background medium, i.e., $\chi_{\rm s}(\mathbf{r},\omega)\neq 0$ only over a finite spatial domain, 
the exponential function in Eq.~\eqref{eq: SinfInt} can be defined as a sequence of functions of the form:
\begin{equation}
    p_\lambda(\alpha)=\exp\left (-2\omega{\rm Im}\left[\sqrt{1+\alpha\chi(\omega)}\right]\lambda/c\right),
\end{equation}
where $\lambda$ is the sequence index. Here, the limits cannot be exchanged, because the sequence of functions $p_\lambda(\alpha)$ does not uniformly converge to $\lim_{\lambda\rightarrow\infty}p_\lambda(\alpha)\equiv p(\alpha)=0$ as a function of $\alpha$ on any interval $\alpha\in[0,r]$ with $r>0$, because
\begin{equation}
    \lim_{\lambda\rightarrow\infty}\underset{\alpha\in[0,r]}{{\rm sup}}|p_\lambda(\alpha)|=\lim_{\lambda\rightarrow\infty}|p_\lambda(0)|=1\neq 0,
\end{equation}
where ``${\rm sup}$'' is the supremum maximum. Importantly, this is also true for the special case of vacuum, i.e., $\chi_{\rm s}(\mathbf{r},\omega)=0$ for all $\mathbf{r}\in\mathbb{R}^3$.

In the case of a lossy spatial-homogeneous medium, i.e., $\chi_{\rm s}(\mathbf{r},\omega)=\chi_{\rm s}(\omega)\neq 0$ for all $\mathbf{r}\in\mathbb{R}^3$,  
 the sequence reads
\begin{equation}
    \tilde{p}_\lambda(\alpha)=\exp(-2\omega{\rm Im}\left[\sqrt{1+\chi_{\rm s}(\omega)+\alpha\chi(\omega)}\right]\lambda/c),
\end{equation}
where $\chi_{\rm s}(\omega)$ is complex. Here,
$\tilde{p}_\lambda(\alpha)$ does uniformly converge to $0$ for any $\alpha\in[0,r]$, since the supremum maximum is still located at $\alpha=0$, but 
\begin{equation}
    \lim_{\lambda\rightarrow\infty}|\tilde{p}_\lambda(0)|=\lim_{\lambda\rightarrow\infty}\exp(-2\omega{\rm Im}\left[\sqrt{1+\chi_{\rm s}(\omega)}\right]\lambda/c)=0.
\end{equation}
The key difference to the former case is, that there is still absorption at $\alpha=0$ ($k_{\alpha=0,I}\neq 0$). The same holds also true in any spatial-inhomogeneous media with a lossy background permittivity $\epsilon_{\rm B}\neq 1$.

In any of the discussed cases, the exponential function and RHS of Eq.~\eqref{eq: SinfInt} vanish.
Since there is no $\lambda$-dependent term left, we can apply the limit $\alpha\rightarrow 0$ on the surviving term from Eq.~\eqref{eq: GFrel2prime} 
to obtain the relation 
\begin{align}
\lim_{\alpha\rightarrow 0}\mathbf{M}_{\alpha}&(\mathbf{r},\mathbf{r}')={\rm Im}\left[\mathbf{G}(\mathbf{r},\mathbf{r}')\right]\label{eq: FDRel1},
\end{align}
which is a corrected version of the relation
\begin{equation}
\lim_{\lambda\rightarrow \infty}\int_{V(\lambda)}{\rm d}\mathbf{r}'' \epsilon_I(\mathbf{r}'') \mathbf{G}(\mathbf{r}, \mathbf{r}'') \cdot \mathbf{G}^*(\mathbf{r}'', \mathbf{r}')=  {\rm Im}\left[\mathbf{G}(\mathbf{r}, \mathbf{r}')\right],\label{eq: GFeps_relation}
\end{equation}
from Ref.~\onlinecite{Dung}.
In fact, if we have started with $\hat{\mathbf{E}}_{\alpha=0}(\mathbf{r},\omega)$
instead of $\hat{\mathbf{E}}_\alpha(\mathbf{r},\omega)$, i.e. setting $\alpha= 0$ in Eq.~\eqref{eq: SinfInt}, the integral over $\mathcal{S}_\infty$ does not vanish for vacuum or finite scattering objects with lossless background medium, although $\mathbf{G}$ vanishes. To achieve this, we look at the vacuum case, $\epsilon(\mathbf{r},\omega)=1$, and thus $\mathbf{G}=\mathbf{G}_{\rm B}$ with $k_{\alpha=0}=\omega/c$. The exponential function on the 
RHS of Eq.~\eqref{eq: SinfInt} would immediately turn to $1$ and we are left with 
\begin{equation}
    \mathbf{I}'_{\rm g}(\mathbf{r},\mathbf{r}')=k_0^4\int_{0}^{2\pi}d\varphi\int_0^{\pi}\sin(\vartheta)d\vartheta e^{-i(\mathbf{r}-\mathbf{r}')\cdot\hat{\mathbf{s}}}(\mathbf{1}-\hat{\mathbf{s}}\hat{\mathbf{s}}),
\end{equation}
where $\hat{\mathbf{s}}$ is the radial basis vector in spherical coordinates. Looking at the special case $\mathbf{r}=\mathbf{r}'=\mathbf{r}_0$, we can perform the angular integrals analytically: 
\begin{equation}
     \mathbf{I}'_{\rm g}(\mathbf{r}_0,\mathbf{r}_0)=-\frac{2i\omega^5}{6c^5\pi}. 
\end{equation}
Putting this back into Eq.~\eqref{eq: GFrel2prime}, with $\alpha=0$ and $\mathbf{G}=\mathbf{G}_{\rm B}$, gives 
\begin{equation}
   M_{\rm B,\alpha=0}(\mathbf{r}_0,\mathbf{r}_0,\omega)={\rm Im}[\mathbf{G}_{\rm B}(\mathbf{r}_0,\mathbf{r}_0)]-\frac{\omega^3}{6\pi c^3}.
\end{equation}

However, we also note that ${\rm Im}[\mathbf{G}_{\rm B}(\mathbf{r}_0,\mathbf{r}_0)]=\omega^3/(6\pi c^3)$ 
and hence $M_{\alpha=0}(\mathbf{r}_0,\mathbf{r}_0,\omega)=0$. This is not consistent with the fluctuation-dissipation theorem, since it implies $\langle0|\hat{\mathbf{E}}^{\dagger}(\mathbf{r}_0,\omega)\hat{\mathbf{E}}(\mathbf{r}_0, \omega')|0\rangle=0$ (cf.~Eq.~\eqref{eq: FDTRelation}), which contradicts with the well known quantum interaction of an atom with free-space environment. In contrast, using the permittivity sequences, we get 
\begin{equation}
    \lim_{\alpha\rightarrow 0}\mathbf{M}_{\rm B,\alpha}(\mathbf{r}_0,\mathbf{r}_0)={\rm Im}\left[\mathbf{G}_{\rm B}(\mathbf{r}_0,\mathbf{r}_0)\right],
\end{equation}
in complete agreement with the fluctuation-dissipation theorem, signified by the relation in Eq.~\eqref{eq: FDTRelation}.
Thus {\it the introduction of $\alpha> 0$ is essential here}.

It follows from Eq.~\eqref{eq: FDRel1} 
that the zero point fluctuations at $\mathbf{r}$ and $\mathbf{r}'$ is equal to the imaginary part of the propagator between points at $\mathbf{r},\mathbf{r}'$, which is a physical appealing result with respect to the fluctuation-dissipation theorem, as discussed in, e.g., Ref.~\onlinecite{grunwel}. However, we already showed that, in the limit of non-absorbing media, the lhs of Eq.~\eqref{eq: GFeps_relation} is zero, and hence the zero-point fluctuations vanish, which means that the formal quantization approach~\cite{Dung,grunwel} cannot describe the limiting case of vacuum fluctuations. This is not the case in Eq.~\eqref{eq: FDRel1} anymore, since the lhs must always be regarded as a limit, where $\lambda \rightarrow \infty$ must be applied before $\alpha\rightarrow 0$.

However, for some application/variations of the Green function quantization approach (e.g., for the QNM quantization below), 
we cannot directly exploit the relation in Eq.~\eqref{eq: FDRel1} and need to calculate a similar form of the LHS of Eq.~\eqref{eq: FDRel1}, which in its current form seems to be impractical, since the integral domain is clearly the whole space. Especially in the important case for applications of a finite nanostructure within a lossless spatial-homogeneous background, it would be desirable to calculate the zero-point fluctuations with integrals over a finite region. To circumvent the integration over a infinite region, and to show that there is a contribution connected to the radiative damping in the 
LHS of Eq.~\eqref{eq: GFeps_relation},
we also derive a different variant of Eq.~\eqref{eq: FDRel1} (and Eq.~\eqref{eq: GFeps_relation}).

We start again with the rhs of Eq.~\eqref{eq: FDTRelation}, but now split $V(\lambda)$ into volume integrals over $V_{\rm in}$ and $V_{\rm out}(\lambda)$.  
Once more we use 
the Helmholtz equation~\eqref{eq: GF_Helmholtz} in combination with the dyadic-dyadic Green second identity, to arrive at (similar to the derivation in App.~\ref{App: ProofMalpha}):
\begin{align}
\mathbf{M}_{\alpha}&(\mathbf{r},\mathbf{r}')=\int_{V_{\rm in}}{\rm d}\mathbf{s}~ \epsilon_I^{(\alpha)}(\mathbf{s})\mathbf{G}_{\alpha}(\mathbf{r}, \mathbf{s}) \cdot \mathbf{G}_{\alpha}^*(\mathbf{s}, \mathbf{r}') \nonumber \\
&+\frac{c^2}{2i\omega^2}\lim_{\lambda\rightarrow\infty}\int_{\mathcal{S}'(\lambda)}{\rm d}A_\mathbf{s}\left\{\mathbf{C}_{\alpha}(\mathbf{s},\mathbf{r},\mathbf{r}')- \mathbf{C}_{\alpha}^\dagger(\mathbf{s},\mathbf{r}',\mathbf{r})\right\}, \label{GFrel2}
\end{align}
with the combined surface $\mathcal{S}'(\lambda)=\mathcal{S}+\mathcal{S}_{\infty}(\lambda)$,
and $V_{\rm in}$ (and its radius $R_{\rm in}$) is chosen, such that $\mathbf{r},\mathbf{r}'\in V_{\rm in}$. 
The integral over $\mathcal{S}_{\infty}$ vanishes as we have already shown earlier.
Since there is no $\lambda$-dependent term left, we can apply the limit $\alpha\rightarrow 0$ on the surviving terms from Eq.~\eqref{GFrel2} (volume integral over $V_{\rm in}$ and surface integral over $\mathcal{S}$), to obtain the relation 
\begin{align}
\lim_{\alpha\rightarrow 0}\mathbf{M}_{\alpha}&(\mathbf{r},\mathbf{r}')=\int_{V_{\rm in}}{\rm d}\mathbf{s}~ \epsilon_{I}(\mathbf{s}) \mathbf{G}(\mathbf{r}, \mathbf{s}) \cdot \mathbf{G}^*(\mathbf{s}, \mathbf{r}') \nonumber \\
&+\frac{c^2}{2i\omega^2}\int_{\mathcal{S}}{\rm d}A_\mathbf{s}~ \Big\{\mathbf{C}(\mathbf{s},\mathbf{r},\mathbf{r}')- \mathbf{C}^\dagger(\mathbf{s},\mathbf{r}',\mathbf{r})\Big\},\label{eq: Malphafin2}
\end{align}
which is one of the main results of this work. 

It is important to note that we can now choose $V_{\rm in}$ and its surrounding surface $\mathcal{S}$ arbitrarily, as long as these remain finite and contain all sources $\epsilon_I(\mathbf{s})$, as well as $\mathbf{r}'$ and $\mathbf{r}$. For a practical evaluation involving finite absorptive nanostructures, the volume integral is always only performed over the absorptive regions. In the case of no absorption, only the surface integral remains. \\

\section{Formal results  of a quasinormal mode quantization scheme\label{Sec: FormQuantQNM}}
\subsection{Quasinormal mode approach\label{Subsec: QNMApproach}}

Here, we recapitulate the definition and properties of QNMs, used for the quantization scheme (described in the next subsection). Similar to the construction of the electric field operator in Subsection~\ref{SubSec: GFquantMod}, we also take into account the permittivity sequences $\epsilon^{(\alpha)}(\mathbf{r},\omega)$. In this context, we define the QNM eigenfunctions $\tilde{\mathbf{f}}_{\mu}(\mathbf{r})$ (implicitly $\alpha$-dependent) with mode index $\mu = \{\pm 1,\pm2 \dots\}$ as solutions to the Helmholtz equation:
\begin{equation}
\boldsymbol{\nabla}\times\boldsymbol{\nabla}\times\tilde{\mathbf{f}}_{\mu}(\mathbf{r})-\frac{\tilde{\omega}_{\mu}^2}{c^2}\epsilon^{(\alpha)}(\mathbf{r},\tilde{\omega}_{\mu})\tilde{\mathbf{f}}_{\mu}(\mathbf{r}) = 0,
\end{equation}
together with outgoing boundary conditions for the specific geometry in Fig.~\ref{fig1}, i.e, the Silver-M\" uller radiation conditions~\cite{KristensenHughes},
\begin{equation}
\frac{\mathbf{r}}{|\mathbf{r}|}\times\boldsymbol{\nabla}\times\tilde{\mathbf{f}}_{\mu}(\mathbf{r})\rightarrow -i\sqrt{\epsilon_{\rm B}^{(\alpha)}(\tilde{\omega}_{\mu})}\frac{\tilde{\omega}_{\mu}}{c}\tilde{\mathbf{f}}_{\mu}(\mathbf{r}),\label{eq: SMcond}
\end{equation}
which are asymptotic conditions for $|\mathbf{r}|\rightarrow \infty$. 
The radiation condition leads to complex QNM eigenvalues $\tilde{\omega}_{\mu}=\omega_{\mu} - i\gamma_{\mu}$, where $\omega_{\mu}$ and $\gamma_{\mu}$ are the center frequency and half width at half maximum or the decay rate of the QNM resonance, respectively. 
The corresponding quality factor is defined as $Q_{\mu}=\omega_{\mu}/(2\gamma_{\mu})$.
We note, that although the QNM eigenfunctions and eigenvalues are implicitly $\alpha$-dependent, $\alpha\chi(\tilde{\omega}_\mu)$ is very small in the frequency interval of interest compared to $\epsilon(\mathbf{r},\tilde{\omega}_\mu)$, if $\omega_0$ is chosen far away from the relevant QNM frequencies (cf. Eq.~\eqref{eq: SmallLorentz}). 

The decay of the QNMs together with the Silver-M\" uller radiation condition leads to a divergent behaviour of the QNM eigenfunctions in far-field regions $V_{\rm out}(\lambda)$. This makes the normalization of these modes a challenging task and it involves  usually volume and surface integrals of the geometry of interest~\cite{MDR1,muljarovPert,SauvanNorm,normaliz}. However, for many practical cases, e.g., for spherical structures with spatial-homogeneous background medium, it was shown~\cite{MDR1} that, for positions inside the scattering geometry, the QNMs form a complete basis or (at least) can be used approximately to expand the full electric field. In particular, the full (transverse) Green function $\mathbf{G}_\alpha(\mathbf{r},\mathbf{r}',\omega)$ can be expanded in these basis functions for positions in the scattering geometry, 
using the representation for the Green dyad~\cite{MDR1,muljarovPert,KristensenHughes,Doost_PRA_87_043827_2013,SauvanNorm},
\begin{equation}
\mathbf{G}_{\rm ff}(\mathbf{r},\mathbf{r}',\omega)=\sum_\mu A_{\mu}(\omega)\tilde{\mathbf{f}}_{\mu}(\mathbf{r})\tilde{\mathbf{f}}_{\mu}(\mathbf{r}'),\label{eq: GFQNM}
\end{equation}
where $\tilde{\mathbf{f}}_{\mu}(\mathbf{r})$ are normalized 
QNMs and the coefficients $A_{\mu}(\omega)$ are given by 
\begin{equation}
A_{\mu}(\omega)=\frac{\omega}{2(\tilde{\omega}_{\mu}-\omega)}.
\end{equation}

We note that there are alternatives forms of $A_{\mu}(\omega)$~\cite{MDR1}, which can be converted into one another by using sum rules of the QNMs, but require additional terms in Eq.~\eqref{eq: GFQNM} (cf.~Ref.~\onlinecite{kristensen2019modeling}). However, we emphasize that the QNM Green function from Eq.~\eqref{eq: GFQNM} with the above choice of $A_\mu(\omega)$ is consistent with the fundamental commutation relation in the quantization approach for lossy and absorptive media from Section~\ref{Sec: Theory} (cf. App.~\ref{app: GFQNMQuant}). We further note, that while the QNM expansion approximates solely the transverse part of the total Green function, the longitudinal part can  easily be obtained from the background Green function (cf. App.~\ref{app: GFQNMQuant}), which however, is 
negligible  in cavity-QED scenarios, where, e.g., a quantum emitter is placed near the scattering object.

To obtain the Green function also for positions outside the resonator geometry (i.e., in the background medium), one can exploit the Dyson equation.
We do this, since the QNMs are not a good representation of the fields
outside the scattering geometry, and they need a
regularization to prevent spatial divergence~\cite{GeNJP2014,RegQNMs}.
Here, we again assume that $\mathbf{G}_{\rm ff}(\mathbf{r},\mathbf{r}',\omega)$ is an accurate approximation to the full Green function inside the resonator geometry region, 
to obtain the Green function for $\mathbf{r},\mathbf{r}'$ outside the scattering geometry as~\cite{GeNJP2014} $\mathbf{G}_{\alpha}(\mathbf{r},\mathbf{r}')=\mathbf{G}_{\rm FF,\alpha}(\mathbf{r},\mathbf{r}')+\mathbf{G}_{\rm others,\alpha}(\mathbf{r},\mathbf{r}')$ with the QNM contribution
\begin{equation}
\mathbf{G}_{\rm FF,\alpha}(\mathbf{r},\mathbf{r}',\omega)=\sum_\mu A_{\mu}(\omega)\tilde{\mathbf{F}}_{\mu}(\mathbf{r},\omega;\alpha)\tilde{\mathbf{F}}_{\mu}(\mathbf{r}',\omega;\alpha),
\end{equation}
and all other contributions associated to the background are summarized in $\mathbf{G}_{\rm others,\alpha}(\mathbf{r},\mathbf{r}')$.
The $\alpha$-dependent fields $\tilde{\mathbf{F}}_{\mu}(\mathbf{r},\omega;\alpha)$ are regularized QNM functions,
\begin{equation}
\tilde{\mathbf{F}}_{\mu}(\mathbf{r},\omega;\alpha)=\int_{V_{\rm in}}{\rm d}\mathbf{r}'\Delta\epsilon^{(\alpha)}(\mathbf{r}',\omega)\mathbf{G}_{\rm B,\alpha}(\mathbf{r},\mathbf{r}',\omega)\cdot\tilde{\mathbf{f}}_{\mu}(\mathbf{r}'),\label{eq: RegF}
\end{equation}
which can be obtained from the QNMs inside the scattering geometry,
or by using numerical calculations of the QNMs in real frequency space~\cite{RegQNMs}. Here, $\Delta\epsilon^{(\alpha)}(\mathbf{r}',\omega)=\epsilon^{(\alpha)}(\mathbf{r}',\omega)-\epsilon_{\rm B}^{(\alpha)}(\omega)$ is the permittivity difference with respect to the background medium and we recall, that $\mathbf{G}_{\rm B,\alpha}(\mathbf{r},\mathbf{r}',\omega)$ is the Green function of the spatial-homogeneous background medium, i.e.,  the solution to the Helmholtz equation Eq.~\eqref{eq: GF_Helmholtz} with $\epsilon(\mathbf{r},\omega)=1$,
together with suitable boundary conditions. We note, that for most practical cases, already a few QNMs are sufficient to accurately approximate the Green function on the range of frequencies of interest~\cite{KamandarDezfouli2017, RegQNMs, PhysRevB.92.205420}. Interestingly, the presence of a discrete set ($\tilde{\mathbf{f}}_\mu(\mathbf{r}))$ and a continuous set ($\tilde{\mathbf{F}}_{\mu}(\mathbf{r},\omega)$) of mode functions occurs also in other modal theories, such as the generalizations of the Carniglia-Mandel modes model~\cite{PhysRevD.3.280}, e.g., the approach in Refs.~\onlinecite{PhysRevA.50.4350,PhysRevA.52.1640}. However, in these approaches, the two set of modes (usually called trapped and traveling modes) are normalized via the standard scalar product (similar to the RHS of Eq.~\eqref{eq: NormNormalM}), restricted to lossless resonators, and both sets are linked via boundary conditions at the dielectric interfaces; while in the QNM model, $\tilde{\mathbf{F}}_{\mu}(\mathbf{r},\omega)$ is obtained from the discrete QNM $\tilde{\mathbf{f}}_\mu(\mathbf{r})$, which itself fulfills outgoing boundary conditions and is normalized via a non-standard bilinear form (cf. Eq.~\eqref{eq: BilinFormQNM}).

In Ref.~\onlinecite{ren_near-field_2020}, it was shown how to construct an approximated regularized QNM via the field equivalence principle~\cite{Neartofar_1992}, where the sources inside the scattering region are replaced by artificial sources on a virtual surface $S'$ around the sources (cf.~Fig.~\ref{fig1}). This yields the same field $\tilde{\mathbf{F}}_{\mu}(\mathbf{r},\omega;\alpha)$ as Eq.~\eqref{eq: RegF} for positions outside the resonator, and within a frequency interval $\Delta\omega$, that covers the resonance of the QNM $\mu$ through: 
\begin{align}
    \tilde{\mathbf{F}}_{\mu}(\mathbf{r},\omega;\alpha)\big\vert_{\Delta\omega}&=i\omega_\mu\mu_0 \oint_{\mathcal{S}'} {\rm d}A_{\mathbf{s}'} \mathbf{G}_{\rm B,\alpha}(\mathbf{r},\mathbf{s}',\omega)\cdot \mathbf{J}_{\rm S'}^{\mu} (\mathbf{s}')\nonumber\\
    &-\oint_{\mathcal{S}'} {\rm d}A_{\mathbf{s}'}\left[\boldsymbol{\nabla}\times \mathbf{G}_{\rm B,\alpha}(\mathbf{r},\mathbf{s}',\omega)\right]\cdot \tilde{\mathbf{M}}_{\rm S'}^{\mu} (\mathbf{s}'), \label{eq: FregFEP}
\end{align}
where 
\begin{align}
\tilde{\mathbf{J}}_{\rm S'}^{\mu}(\mathbf{s}')=\mathbf{n}_{\mathbf{s}}'\times\tilde{\bf{h}}_{\mu}(\mathbf{s}'),\\
\tilde{\mathbf{M}}_{\rm S'}^{\mu}(\mathbf{s}')=-\mathbf{n}_{\mathbf{s}}'\times\tilde{\bf{f}}_{\mu}(\mathbf{s}'),
\end{align}
are the sources on the boundary,
and $\tilde{\mathbf{h}}_{\mu}(\mathbf{r}')=\boldsymbol{\nabla}\times\mathbf{\tilde{f}_{\mu}(r')}/(i\tilde{\omega}_{\mu}\mu_{0})$ is the magnetic field of the associated QNM $\mu$ and $\mathbf{n}_{\mathbf{s}}'$ is the normal vector on the surface $\mathcal{S}'$ and pointing outwards with respect to the enclosed volume.

Note that indeed within a small frequency interval around $\omega_\mu$, where $\epsilon(\mathbf{r},\omega)$ is approximately constant and $Q_\mu>10$, the regularized expression from Eq.~\eqref{eq: RegF} and  Eq.~\eqref{eq: FregFEP} 
are found to be practically the same in the far field (cf.~Ref.~\onlinecite{ren_near-field_2020}). We also highlight
that the approximate expression from the near field to far field transformation can simplify the numerical calculations involving, e.g., far field propagation of QNMs, significantly compared to Eq.~\eqref{eq: RegF}, since the internal volume integral is replaced by a surface integral
(for numerical details, cf. Ref.~\onlinecite{ren_near-field_2020}). 
However, for single mode behaviour, one
can also adopt a highly accurate regularization procedure~\cite{RegQNMs}
to obtain the field inside and outside the resonator,
as we will discuss below for the numerical examples. 

\subsection{Quasinormal mode quantization scheme}
Next, we generalize the QNM quantization procedure from Ref.~\onlinecite{PhysRevLett.122.213901} using permittivity sequences. 
Using the  approximated form of the full Green function using QNMs, i.e., $\mathbf{G}_{\rm ff,\alpha}(\mathbf{r},\mathbf{r}',\omega)$ and the regularization $\mathbf{G}_{\rm FF,\alpha}(\mathbf{r},\mathbf{r}',\omega)$, 
we can rewrite the electric field operator from Eq.~\eqref{eq: FormSolE} in terms of QNMs.  
For positions $\mathbf{r}$ nearby the scattering geometry, we obtain the expression for the total electric field operator $\hat{\mathbf{E}}_\alpha(\mathbf{r})=\int_0^{\infty}{\rm d}\omega \hat{\mathbf{E}}_\alpha(\mathbf{r},\omega)+{\rm H.a.}$ (from Eq.~\eqref{eq: FormSolE}), 
\begin{equation}
\hat{\mathbf{E}}_\alpha(\mathbf{r})=i\sum_\mu\sqrt{\frac{\omega_\mu}{2\epsilon_0}}\tilde{\mathbf{f}}_\mu(\mathbf{r})\tilde{a}_\mu +{\rm H.a.}, \label{eq: EQNMfieldUnsym}
\end{equation}
where 
\begin{align}
\tilde{a}_\mu = \int_{0}^\infty\frac{\sqrt{2}A_\mu(\omega)}{\sqrt{\pi\omega_\mu}}\Bigg(&\int_{V_{\rm in}}{\rm d}\mathbf{r}\sqrt{\epsilon_I^{(\alpha)}(\mathbf{r},\omega)}\tilde{\mathbf{f}}_\mu(\mathbf{r})\cdot\hat{\mathbf{b}}_\alpha(\mathbf{r},\omega)\nonumber\\
&+\lim_{\lambda\rightarrow \infty}\int_{V_{\rm out}(\lambda)}{\rm d}\mathbf{r}\sqrt{\epsilon_I^{(\alpha)}(\mathbf{r},\omega)}\nonumber\\
&\times\tilde{\mathbf{F}}_{\mu}(\mathbf{r},\omega;\alpha)\cdot\hat{\mathbf{b}}_{\alpha}(\mathbf{r},\omega)\Bigg){\rm d}\omega, \label{eq: QNMOperator}
\end{align}
is a QNM operator, that depends implicitly on $\alpha$.
 
The first term corresponds to the medium  in the absorptive regions ($V_{\rm in}$), while the second part corresponds to the contribution from the photons.

As shown in  Ref.~\onlinecite{PhysRevLett.122.213901}, applying a symmetrization transformation to the operators $\tilde{a}_\mu(\tilde{a}_\mu^\dagger)$,
we obtain 
\begin{equation}
\hat{\mathbf{E}}(\mathbf{r})=i\sum_\mu\sqrt{\frac{\omega_\mu}{2\epsilon_0}}\tilde{\mathbf{f}}_\mu^s(\mathbf{r})a_\mu +{\rm H.a.},
\end{equation}
where $a_\mu (a_\mu^{\dagger})$ are suitable annihilation (creation) operators acting on the symmetrized QNM Fock subspace and $\tilde{\mathbf{f}}_\mu^{\rm s}(\mathbf{r})=\sum_\eta\sqrt{\omega_\eta/\omega_\mu}\left(\mathbf{S}^{1/2}\right)_{\eta\mu}\tilde{\mathbf{f}}_\eta(\mathbf{r})$ are the symmetrized QNM eigenfunctions. The 
radiative and material losses
give rise to a loss induced symmetrization matrix,
with matrix elements:
\begin{equation}
S_{\mu\eta,\alpha} = \int_{0}^\infty \frac{2A_\mu(\omega)A_{\eta}^*(\omega)}{\pi\sqrt{\omega_\mu\omega_\eta}}\left(\tilde{S}^{\rm in}_{\mu\eta,\alpha}(\omega)+\tilde{S}^{\rm out}_{\mu\eta,\alpha}(\omega)\right)\label{eq: Sfull},
\end{equation}
where 
\begin{equation}
\tilde{S}^{\rm in}_{\mu\eta,\alpha}(\omega) = \int_{V_{\rm in}} {\rm d}\mathbf{r}\epsilon_I^{(\alpha)}(\mathbf{r},\omega)\tilde{\mathbf{f}}_{\mu}(\mathbf{r})\cdot\tilde{\mathbf{f}}_{\eta}^*(\mathbf{r}),\label{eq: Sin}
\end{equation}
and 
\begin{align}
&\tilde{S}^{\rm out}_{\mu\eta,\alpha}(\omega)\nonumber \\
&= \lim_{\lambda\rightarrow \infty}\int_{V_{\rm out}(\lambda)} {\rm d}\mathbf{r}\epsilon_I^{(\alpha)}(\mathbf{r},\omega)\tilde{\mathbf{F}}_{\mu}(\mathbf{r},\omega;\alpha)\cdot\tilde{\mathbf{F}}_{\eta}^*(\mathbf{r},\omega;\alpha).\label{eq: Sout}
\end{align}

At first glance, it seems that in the non-absorptive limit of $\epsilon_I\rightarrow 0$, would yield $S_{\mu\eta}=0$, and hence all QNM operator related quantities vanish. However, as explained earlier, the exchange of the limits $\alpha\rightarrow 0$ and $\lambda\rightarrow \infty$ is non-trivial. As we have shown in Section~\ref{Sec: FDTheorem}, there is a general relation between fluctuation and dissipation including a boundary term, and with the help of the permittivity sequences, we will subsequently resolve this 
apparent issue in Section~\ref{Sec: Commutation}.

\section{Commutation relation of the QNM operators in the dielectric limit\label{Sec: Commutation}}

Having discussed the general relation between fluctuation and dissipation, we will now apply the ideas from Section~\ref{Sec: FDTheorem} to the QNM quantization scheme from Section~\ref{Sec: FormQuantQNM}.

Inspecting Eqs.~\eqref{eq: Sfull}-\eqref{eq: Sout} suggests that $\lim_{\alpha\rightarrow 0}S_{\mu\eta,\alpha}$ seem to vanish in the limit of non-absorbing media, i.e., when $\epsilon_I \rightarrow 0$. In the following, we 
show explicitly that $S_{\mu\eta}$ does not vanish in the limit without absorption, e.g., in the case of a dielectric cavity structure embedded in vacuum.
 In the non-absorptive limit, obviously $\tilde{S}^{\rm in}_{\mu\eta,\alpha}(\omega)$ vanishes immediately, since it is independent of $\lambda$. This contribution reflects the absorptive part of the mode dissipation. Furthermore, we note that the expression for $\tilde{S}^{\rm in}_{\mu\eta,\alpha}(\omega)$ for a single mode is similar to the classical absorptive part connected to the Poynting theorem~\cite{liberal2019control}, and we will henceforth call it the non-radiative part $\tilde{S}^{\rm in}_{\mu\eta,\alpha}(\omega)=\tilde{S}^{\rm nrad}_{\mu\eta,\alpha}(\omega)$. We are left with the contribution connected to $\tilde{S}^{\rm out}_{\mu\eta}(\omega)=\lim_{\alpha\rightarrow 0}\lim_{\lambda\rightarrow \infty}\tilde{S}^{\rm out}_{\mu\eta,\alpha\lambda}(\omega)$:
\begin{align}
&\tilde{S}^{\rm out}_{\mu\eta,\alpha\lambda}(\omega)=\int_{V_{\rm out}(\lambda)} {\rm d} \mathbf{r}~\epsilon_{I}^{(\alpha)} (\mathbf{r},\omega)\tilde{\mathbf{F}}_{\mu}(\mathbf{r},\omega;\alpha)\cdot\tilde{\mathbf{F}}_{\eta}^*(\mathbf{r},\omega;\alpha).\label{eq: Sout_alpha}
\end{align}

Next we discuss and compare different approaches to obtain a non-vanishing contribution for $\tilde{S}^{\rm out}_{\mu\eta,\alpha\lambda}(\omega)$ in the limits $\lambda\rightarrow \infty$ and $\alpha\rightarrow 0$ applied in this
order, which will yield the second term of the Poynting theorem, reflecting radiative loss.

\subsection{Helmholtz equation and Green second identity}
Here we  apply directly a  method similar to that described in Section~\ref{Sec: FDTheorem}. We  use the regularized modes $\tilde{\mathbf{F}}_{\mu}(\mathbf{r},\omega;\alpha)$ from Eq.~\eqref{eq: RegF}, and we note that the derivation is analogue to the case where $\tilde{\mathbf{F}}_{\mu}(\mathbf{r},\omega;\alpha)$ is obtained from the field equivalence principle (cf. Eq.~\eqref{eq: FregFEP}). We first use the definition of the regularized QNM fields $\tilde{\mathbf{F}}_{\mu}(\mathbf{r},\omega;\alpha)$ from Eq.~\eqref{eq: RegF} and Eq.~\eqref{eq: Sout_alpha} to obtain 
\begin{align}
\tilde{S}^{\rm out}_{\mu\eta,\alpha\lambda}(\omega) =& \int_{V_{\rm in}}{\rm d}\mathbf{r}'\int_{V_{\rm in}}{\rm d}\mathbf{r}''\Delta\epsilon^{(\alpha)}(\mathbf{r}')\Delta\epsilon^{(\alpha)*}(\mathbf{r}'')\nonumber\\
&\times\tilde{\mathbf{f}}_{\mu}(\mathbf{r}')\cdot\mathbf{M}_{\alpha,\lambda}^{\rm B}(\mathbf{r}',\mathbf{r}'')\cdot\tilde{\mathbf{f}}_{\eta}^*(\mathbf{r}'')\label{eq: Soutalpha2},
\end{align}
with 
\begin{equation}
\mathbf{M}_{\alpha,\lambda}^{\rm B}(\mathbf{r}',\mathbf{r}'')=\int_{V_{\rm out}(\lambda)} {\rm d}\mathbf{r}\epsilon_I^{(\alpha)}(\mathbf{r})\mathbf{G}_{{\rm B},\alpha}(\mathbf{r}',\mathbf{r})\cdot\mathbf{G}_{{\rm B},\alpha}^*(\mathbf{r},\mathbf{r}''),
\end{equation}
where   $\omega$  is implicitly included (to simplify the notation). 

We remark that we can apply the limit $\alpha\rightarrow 0$ immediately to most parts on the rhs of Eq.~\eqref{eq: Soutalpha2}, except for $\mathbf{M}_{\alpha,\lambda}^{\rm B}(\mathbf{r}',\mathbf{r}'')$, such that the volume integrals over $V_{\rm in}$ reduces to the scattering region inside $V_{\rm in}$. Now using $\epsilon_I=(\epsilon-\epsilon^*)/(2i)$, and noting that $\epsilon^{(\alpha)}(\mathbf{r})=\epsilon_{{\rm B}}^{(\alpha)}$ for $\mathbf{r}\in V_{\rm out}(\lambda)$, we can exploit the Helmholtz equation of the background Green function from (Eq.~\eqref{eq: GF_Helmholtz} with $\epsilon(\mathbf{r},\omega)=1$), together with the dyadic-dyadic Green second identity (Eq.~\eqref{eq: DyadDyadGF}), to arrive at
\begin{align}
\mathbf{M}_{\alpha,\lambda}^{\rm B}&(\mathbf{r}',\mathbf{r}'') \nonumber \\
=&\frac{c^2}{2i\omega^2}\oint_{\mathcal{S}'(\lambda)} {\rm d}A_\mathbf{s}\left\{\mathbf{C}_{\rm B,\alpha}(\mathbf{s},\mathbf{r}',\mathbf{r}'')- \mathbf{C}_{\rm B,\alpha}^\dagger(\mathbf{s},\mathbf{r}'',\mathbf{r}')\right\}\label{eq: RelR},
\end{align}
with
\begin{equation}
\mathbf{C}_{\rm B,\alpha}(\mathbf{s},\mathbf{r}',\mathbf{r}'')=\left[\mathbf{n}_\mathbf{s}\times\mathbf{G}_{\rm B,\alpha}(\mathbf{s},\mathbf{r}')\right]^{\rm T} \cdot \left[\boldsymbol{\nabla}_{\mathbf{s}}\times\mathbf{G}_{\rm B,\alpha}^*(\mathbf{s}, \mathbf{r}'')\right]\label{eq: CBdyad}.
\end{equation}
Here, $\mathbf{n}_\mathbf{s}$ is the normal vector of the surface $S+S_{\infty}(\lambda)$ pointing outwards with respect to $V_{\rm out}$.
Equation~\eqref{eq: RelR} is a special case of Eq.~\eqref{GFrel2}, since $\mathbf{r}', \mathbf{r}''$ are always inside the scattering region, by construction of the regularized QNM fields $\tilde{\mathbf{F}}_{\mu}(\mathbf{r},\omega;\alpha)$. Following a similar derivation as in Section~\ref{Sec: FDTheorem}, we get in the limit $\lambda\rightarrow \infty$ the remaining contribution 
\begin{align}
    \lim_{\lambda\rightarrow\infty}&\mathbf{M}_{\alpha,\lambda}^{\rm B}(\mathbf{r}',\mathbf{r}'')\nonumber \\
    &=\frac{c^2}{2i\omega^2}\oint_{\mathcal{S}} {\rm d}A_\mathbf{s}\left\{\mathbf{C}_{\rm B,\alpha}(\mathbf{s},\mathbf{r}',\mathbf{r}'')- \mathbf{C}_{\rm B,\alpha}^\dagger(\mathbf{s},\mathbf{r}'',\mathbf{r}')\right\}.
\end{align}

Taking the limit $\lambda\rightarrow \infty$ in Eq.~\eqref{eq: Soutalpha2}, and inserting  $\lim_{\lambda\rightarrow\infty}\mathbf{M}_{\alpha,\lambda}^{\rm B}(\mathbf{r}',\mathbf{r}'')$ into the resulting equation yields $\tilde{S}^{\rm out}_{\mu\eta,\alpha}(\omega)\equiv\lim_{\lambda\rightarrow \infty}\tilde{S}^{\rm out}_{\mu\eta,\alpha\lambda}(\omega)
$ with
\begin{align}
&\tilde{S}^{\rm out}_{\mu\eta,\alpha}(\omega)\nonumber \\
=& \frac{c^2}{2i\omega^2}\int_{V_{\rm in}}{\rm d}\mathbf{r}'\int_{V_{\rm in}}{\rm d}\mathbf{r}''\Delta\epsilon^{(\alpha)}(\mathbf{r}')\Delta\epsilon^{(\alpha)*}(\mathbf{r}'')\tilde{\mathbf{f}}_{\mu}(\mathbf{r}')\nonumber\\
&\times\left[\oint_{\mathcal{S}} {\rm d}A_\mathbf{s}\left\{\mathbf{C}_{\rm B,\alpha}(\mathbf{s},\mathbf{r}',\mathbf{r}'')- \mathbf{C}_{\rm B,\alpha}^\dagger(\mathbf{s},\mathbf{r}'',\mathbf{r}')\right\}\right]\cdot\tilde{\mathbf{f}}_{\eta}^*(\mathbf{r}'') .\label{eq: Soutalpha3}
\end{align}

Using the definition of $\mathbf{C}_{\rm B,\alpha}(\mathbf{s},\mathbf{r}',\mathbf{r}'')$ from Eq.~\eqref{eq: CBdyad}, and exploiting the definition of $\tilde{\mathbf{F}}_{\mu}(\mathbf{r},\omega;\alpha)$,
we obtain
\begin{align}
\tilde{S}^{\rm out}_{\mu\eta,\alpha}(\omega) =&\frac{ic^2}{2\omega^2}\oint_{\mathcal{S}} {\rm d}A_\mathbf{s}\left[\mathbf{n}_{\mathbf{s}}\times\left(\boldsymbol{\nabla}\times\tilde{\mathbf{F}}_{\mu}(\mathbf{s};\alpha)\right)\right]\cdot\tilde{\mathbf{F}}_{\eta}^*(\mathbf{s};\alpha)\nonumber\\
&-{\rm h.c.}(\mu\leftrightarrow\eta)\label{eq: Soutalpha4},
\end{align}
which does not vanish in the limit $\alpha\rightarrow 0 $ and hence $\epsilon_I\rightarrow 0$. 
Thus, after taking the limit $\alpha\rightarrow 0$, we arrive at $\tilde{S}^{\rm rad}_{\mu\eta}(\omega)\equiv \lim_{\alpha\rightarrow 0}\tilde{S}^{\rm out}_{\mu\eta,\alpha}(\omega)$,
\begin{align}
    &\tilde{S}^{\rm rad}_{\mu\eta}(\omega) \nonumber \\
    &=\frac{1}{2\omega\epsilon_0}\oint_{\mathcal{S}} {\rm d}A_\mathbf{s}\left[\tilde{\mathbf{H}}_{\mu}(\mathbf{s})\times\mathbf{n}_{\mathbf{s}}\right]\cdot\tilde{\mathbf{F}}_{\eta}^*(\mathbf{s})+{\rm h.c.}(\mu\leftrightarrow\eta)\label{eq: Soutfinomega},
\end{align}
where $\tilde{\mathbf{H}}_{\mu}(\mathbf{s})=\boldsymbol{\nabla}\times\tilde{\mathbf{F}}_{\mu}(\mathbf{s})/(i\omega\mu_0)$ is the regularized QNM magnetic field and we note, that $\mathbf{n}_{\mathbf{s}}$ points inwards of $\mathcal{S}$ with respect to $V_{\rm in}$. 
Note that for a single QNM the expression in Eq.~\eqref{eq: Soutfinomega} is indeed similar to the classical radiative output flow connected to the Poynting theorem~\cite{liberal2019control,Hughes_SPS_2019}. 

We remark that the surface $\mathcal{S}$ can be chosen arbitrarily as long as it is far away from the resonator region, such that $\mathbf{G}_{\rm FF}$ is an accurate approximation to the full Green function expansion outside the resonator.

In the limit of no radiative
and no nonradiative loss, namely in the
non-dissipative limit of $\gamma_\mu=0$,
in App.~\ref{app: JCmodelBreakdown2}
we show that
$S_{\mu \mu} \rightarrow 1$.
Thus the model  fully recovers the 
well known result for normal mode
quantization~\cite{GirishBook1}.

\subsection{Far field expression of $\tilde{S}^{\rm rad}_{\mu\eta}(\omega)$}

Equation \eqref{eq: Soutfinomega} is already a significant result and confirms the derivations in Ref.~\onlinecite{PhysRevLett.122.213901}, namely  that there is
indeed a non-vanishing contribution in the case of non-absorptive cavities. However, it would be desirable to find an expression which only depends on integrals and quantities in the system region. In the following, we will use the fact that the approximations involving $\tilde{\mathbf{F}}_{\mu}(\mathbf{s})$ as a function of position outside the scattering region improves with increased distance from the resonator. We will then derive an exact limit of $\tilde{S}^{\rm rad}_{\mu\eta}(\omega)$. 

First, we choose $\mathcal{S}=\mathcal{S}_{\rm far}$ as a spherical surface in the far field region, such that $\mathbf{n}_{\mathbf{s}}=-\mathbf{s}/|\mathbf{s}|$. Next, we recapitulate the radiation condition of the background Green function:
\begin{equation}
\frac{\mathbf{s}}{|\mathbf{s}|}\times\boldsymbol{\nabla}_{\mathbf{s}}\times\mathbf{G}_{\rm B}(\mathbf{s},\mathbf{r})\rightarrow -i\sqrt{\epsilon_{\rm B}(\omega)}\frac{\omega}{c}\mathbf{G}_{\rm B}(\mathbf{s},\mathbf{r})\label{eq: SMcondGFB},
\end{equation}
for $|\mathbf{s}|\rightarrow\infty$, which is an analogue of the Silver-M\"uller radiation condition from Eq.~\eqref{eq: SMcond} for real frequencies $\omega$. Since the regularized QNM obtained from the Dyson approach as well as the field equivalence principle, involve only terms proportional to integrals over $\mathbf{G}_{\rm B}(\mathbf{s},\mathbf{r})$, we use this relation and apply it on the far field surface $\mathcal{S}_{\rm far}$ (Eq.~\eqref{eq: Soutfinomega}), to obtain 
\begin{align}
    \tilde{S}^{\rm rad}_{\mu\eta}(\omega)= \frac{\sqrt{\epsilon_{\rm B}(\omega)}c}{\omega}\oint_{\mathcal{S}_{\rm far}} {\rm d}A_\mathbf{s}\tilde{\mathbf{F}}_{\mu}(\mathbf{s})\cdot\tilde{\mathbf{F}}_{\eta}^*(\mathbf{s})\label{eq: Soutfinomega2}.
\end{align}

Since $\mathcal{S}_{\rm far}$ is located in the far field $|\mathbf{s}|\gg |\mathbf{r}|$ and only the far-field contributions of $\mathbf{G}_{\rm B}$ play a significant role, we can rewrite $\tilde{S}^{\rm rad}_{\mu\eta}(\omega)$ by using the limiting case of $\mathbf{G}_{\rm B}$ for far field positions in Eq.~\eqref{eq: K1func} and \eqref{eq: K2func} to get 
\begin{equation}
    \tilde{S}^{\rm rad}_{\mu\eta}(\omega)= \frac{\sqrt{\epsilon_{\rm B}(\omega)}c}{\omega}\frac{1}{16\pi^2}\int_{0}^{2\pi}d\varphi\int_0^{\pi}\sin(\vartheta) d\vartheta I_{\mu\eta}(\omega), \label{eq: Soutfin3}
\end{equation}
with the two representations $I_{\mu\eta}(\omega)= I^{\rm vol}_{\mu\eta}(\omega)= I^{\rm sur}_{\mu\eta}(\omega)$, yielding
\begin{align}
    I^{\rm vol}_{\mu\eta}(\omega)=&\left(\int_V \Delta\epsilon(\mathbf{r})\mathbf{K}^{(1)}(\hat{\mathbf{s}},\mathbf{r})\cdot\tilde{\mathbf{f}}_\mu(\mathbf{r})\right)\nonumber\\
    &\times\left(\int_V \Delta\epsilon^*(\mathbf{r})\mathbf{K}^{(1)*}(\hat{\mathbf{s}},\mathbf{r})\cdot\tilde{\mathbf{f}}_\eta^*(\mathbf{r})\right), 
    \end{align}
using the Dyson approach (Eq.~\eqref{eq: RegF}), and 
\begin{align}\label{I_sur}
    &I^{\rm sur}_{\mu\eta}(\omega)
    \nonumber\\
    &=\left[\int_{S'} {\rm d}A_{\mathbf{s}'}\left[ \mathbf{K}^{(1)}(\hat{\mathbf{s}},\mathbf{s}')\mathbf{J}_\mu(\mathbf{s}')-\mathbf{K}^{(2)}(\hat{\mathbf{s}},\mathbf{s}')\mathbf{M}_\mu(\mathbf{s}')\right]\right)
     \nonumber\\
    &\times\left[\int_{S'} {\rm d}A_{\mathbf{s}'}\left[ \mathbf{K}^{(1)*}(\hat{\mathbf{s}},\mathbf{s}')\mathbf{J}_\eta^*(\mathbf{s}')-\mathbf{K}^{(2)*}(\hat{\mathbf{s}},\mathbf{s}')\mathbf{M}_\eta^*(\mathbf{s}')\right]\right], 
\end{align}
using the field equivalence principle (Eq.~\eqref{eq: FregFEP}). 

The dyadic forms $\mathbf{K}^{(i)}(\hat{\mathbf{s}},\mathbf{r})$ are given implicitly in Eq.~\eqref{eq: K1func} and \eqref{eq: K2func}. We see that Eq.~\eqref{eq: Soutfin3} has no explicit appearance of the far field surface and only involves integrals over the system regions, which refines the QNM quantization;  $\tilde{S}_{\mu\eta}^{\rm rad/nrad}(\omega)$
and all related quantities in the QNM quantum model can be calculated from the QNMs within the system of interest and all integrals are performed over the scattering region or bounded surfaces. In addition, it was recently pointed out that relations such as in Eq.~\eqref{eq: Soutfinomega} are generally not positive definite, because, e.g., of optical backflow~\cite{wiersig2019nonorthogonality}. However, the (exact) far field limit explicitly shows, that $S_{\mu\eta}^{\rm rad}$ is a positive definite form, since it constitutes a scalar product between different entries $\mu$ and $\eta$. Indeed, one has to choose the surface $\mathcal{S}$ in the general expression from Eq.~\eqref{eq: Soutfinomega} sufficiently far away from the resonator, where $\tilde{\mathbf{F}}_\mu$ is a suitable representation of the fields. In fact, as shown in App.~\ref{App: NumPCBeam}, one needs to choose the integration surface in the general expression, Eq.~\eqref{eq: Soutfinomega}, in the intermediate to far-field region with respect to the resonator in order to achieve numerical convergence.

\subsection{ Numerical example for  three-dimensional photonic crystal beam cavities\label{Subsec: Example}}

\begin{figure}
    \centering
    \includegraphics[width=0.99\columnwidth]{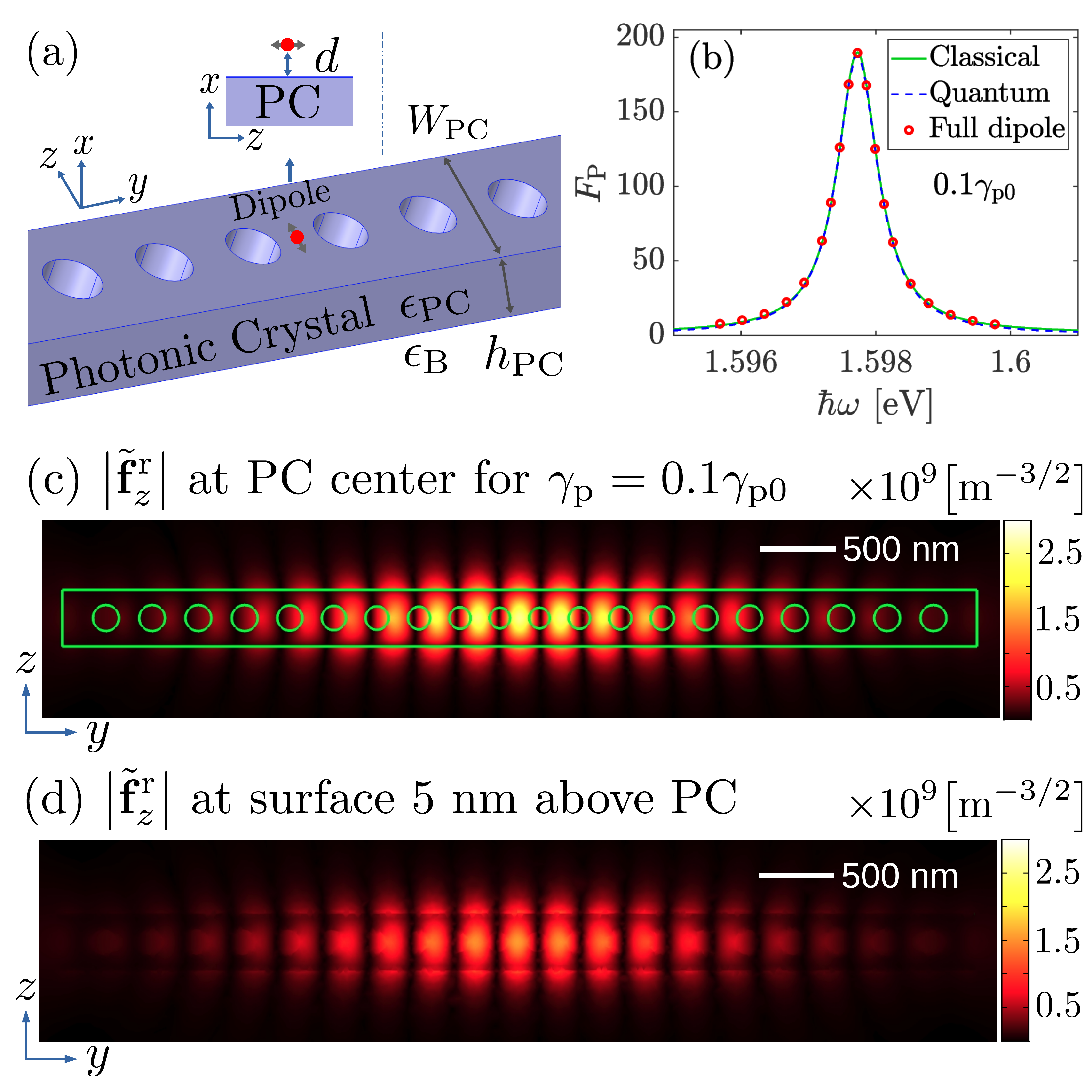}
\caption{(a) Schematic diagram of PC beam with permittivity $\epsilon_{\rm PC}$ in free space ($\epsilon_{\rm B}=1.0$). The $z$-polarized dipole is put at $d=5$ nm above the PC cavity. The width and the height of the PC beam is $W_{\rm PC}=376$ nm and $h_{\rm PC}=200$ nm. (b) Classical Purcell factor $F_{\rm P}^{\rm QNM}$ (Eq.~\eqref{QNMpurcell}) from single QNM, quantum Purcell factor $F_{\rm P}^{\rm quan}$ (Eq.~\eqref{quantumpurcellsingle}) from quantized QNMs, and numerical Purcell factor $F_{\rm P}^{\rm num}$ (Eq.~\eqref{Purcellfulldipole}) from full dipole method for $\gamma_{\rm p}=0.1\gamma_{\rm p0}$. Regularized QNM field $|\tilde{\mathbf{f}}^{\rm r}_{z}|$ (z-component) (c) at PC center surface and (d) at a surface $5$ nm above PC surface for $\gamma_{\rm p}=0.1\gamma_{\rm p0}$. The QNM calculations with different losses are not shown, as they look similar.}
    \label{pc_sche}
\end{figure}

Next, we discuss a practical numerical example 
to study the QNM quantization 
as a function of material loss and 
also in
the limit $\epsilon_I\rightarrow 0$, using a 
photonic crystal (PC) beam cavity (see Fig.~\ref{pc_sche} (a)), whose real part of the
dielectric constant is similar to silicon nitride \cite{KamandarDezfouli2017}. The three-dimensional beam supports
a single QNM with index `c' ({\em cavity} mode) in the spectral regime of interest, which can give rise to a large Purcell factor.

The precise details of the structure and numerical implementation details are given in App.~\ref{App: NumPCBeam}. Here we just mention that
the QNM function $\tilde{\mathbf{f}}^{\rm }_{\rm c}$ and complex frequency $\tilde{\omega}_{\rm c}=\omega_{\rm c}-i\gamma_{\rm c}$ 
are calculated using the method from Ref.~\onlinecite{Bai} in combination with the normalization method shown in Ref.~\onlinecite{RegQNMs}.
In addition,
because of our chosen numerical implementation, these modes are regularized QNMs ($\tilde{{\bf f}}^{\rm r}$), 
computed in real frequency space, so they have the correct behaviour in the far field, but we refer to them here as just the QNM
for ease of notation. In the near field regime,
then $\tilde {\bf f}^{\rm r} \rightarrow \tilde {\bf f}^{\rm }$,
while in the far field regime,
$\tilde {{\bf f}}^{\rm r}\rightarrow \tilde {\bf F}^{\rm }$.
Thus,  
we construct the Green function through~\cite{LeungSP1D,GeNJP2014}
\begin{equation}
\mathbf{G}\left(\mathbf{r},\mathbf{r}_{0},\omega\right)=A_{\rm c}\left(\omega\right)\,\tilde{\mathbf{f}}^{\rm r}_{\rm c}\left(\mathbf{r}\right)\tilde{\mathbf{f}}_{\rm c}^{\rm r}\left(\mathbf{r}_{0}\right),\label{eq:GFwithSUM}
\end{equation}
with $A_{\rm c}(\omega)=\omega/[2(\tilde{\omega}_{\rm c}-\omega)]$ for all spatial locations and at frequencies close to the resonance frequency $\omega_{\rm c}$~\cite{RegQNMs}.

The permittivity $\epsilon_{\rm PC}(\omega)$ of the PC beam is described by a single Lorentz oscillator model,  
\begin{equation}\label{Lorentz}
\epsilon_{\rm PC}(\omega)=\epsilon_{\infty}-\frac{(\epsilon_{\rm s}-\epsilon_{\infty})\omega_{\rm t}^{2}}{\omega^{2}-\omega_{\rm t}^{2}+i\omega\gamma_{\rm p}},
\end{equation}
where $\epsilon_{\infty}=1.0$, $\epsilon_{\rm s}=2.04^2=4.1616$. A relatively large Lorentz resonance energy $\hbar\omega_{\rm t}=12$ eV is used to ensure that it is far away from the PC resonance. We also use $\hbar\gamma_{\rm p}=\hbar\gamma_{\rm p0}=0.131$ eV, and  can easily cover low to high loss regions with the same model by changing the loss term; below we use  $\gamma_{\rm p}$ 
values of $0.5\gamma_{\rm p0}$, $0.3\gamma_{\rm p0}$, $0.2\gamma_{\rm p0}$, and $0.1\gamma_{\rm p0}$ and $\gamma_{\rm p}\rightarrow 0$ to analyze the limit behaviour of $S_{\rm cc}^{\rm rad}\equiv S^{\rm rad}$, $S_{\rm cc}^{\rm nrad}\equiv S^{\rm nrad}$ and $S_{\rm cc}=S$, from finite to vanishing absorption. 
Outside the PC beam, we consider free space with permittivity $\epsilon_{\rm B}=1.0$. The $z$-polarized dipole is 
positioned at $d=5$ nm above the PC cavity (Fig.~\ref{pc_sche} (a)).

The classical generalized Purcell factor for a emitter with dipole moment $\mathbf{d}$ (=d$\mathbf{n}_{\rm d}$), at location $\mathbf{r}_{0}$,
can be obtained though Green function via~\cite{Anger2006,NormKristHughes} 
 \begin{align}\label{QNMpurcell}
 \begin{split}
     F_{{\rm P}}^{\rm QNM}({\bf r}_0,\omega) =1+\frac{6\pi c^{3}}{\omega^{3}n_{\rm B}}\,\mathbf{n}_{\rm d}\cdot{\rm Im}\{\mathbf{G}\left(\mathbf{r}_0,\mathbf{r}_{0},\omega\right)\}\cdot\mathbf{n}_{\rm d},
 \end{split}
 \end{align}
where $n_{\rm B}$ (=$\sqrt{\epsilon_{\rm B}}=1$) is the background refractive index,
and the emitter is assumed to be outside the beam which accounts for the extra factor of 1~\cite{GeNJP2014}.

It is also useful to 
compare with full dipole solutions to Maxwell's equations
(i.e., with no approximations) to check the accuracy of the QNM
expansion and to confirm that we only need one QNM. 
The numerical Purcell factor is defined as follows
\begin{equation}\label{Purcellfulldipole}
    F_{\rm P}^{\rm num}(\mathbf{r}_{0},\omega)=\frac{\int_{ \mathcal{S}_{\rm dipole}}\mathbf{n}_\mathbf{r}\cdot {\bf S}_{\rm dipole,total}(\mathbf{r},\omega){\rm d}{A_\mathbf{r}} }{\int_{\mathcal{S}_{\rm  dipole}}\mathbf{n}_\mathbf{r}\cdot {\bf S}_{\rm dipole,background}(\mathbf{r},\omega){\rm d}{A_\mathbf{r}} },
\end{equation}
where $\mathcal{S}_{\rm dipole}$ is a small spherical surface (with radius $1$ nm) surrounding the dipole point and $\mathbf{n}_{\mathbf{r}}$ is a unit vector normal to $\mathcal{S}_{\rm dipole}$, pointing outward.
The vector ${\bf S}(\mathbf{r},\omega)$ is the Poynting vector at this small surface and the subscript terms `total' and `background' represent the case with and without resonator.

As shown in Fig.~\ref{pc_sche} (b), there is an excellent agreement between the classical Purcell factor $F_{\rm P}^{\rm QNM}$ (Eq.~\eqref{QNMpurcell}) using single QNM theory and the full-dipole numerical Purcell factor $F_{\rm P}^{\rm num}$ (Eq.~\eqref{Purcellfulldipole}); here we use  $\gamma_{\rm p}=0.1\gamma_{\rm p0}$, but 
more comparisons are shown explicitly in App.~\ref{App: NumPCBeam} for different loss values.
In addition, the numerical radiative and nonradiative beta factors are defined as 
\begin{gather}
    \beta_{\rm num}^{\rm rad}(\mathbf{r}_{0},\omega)=\frac{\int_{\mathcal{S}_{\rm PML}}\mathbf{n}_{\mathbf{r}_{\rm PML}}\cdot {\bf S}_{\rm PML,total}(\mathbf{r}_{\rm PML},\omega){\rm d}{A_{\mathbf{r}_{\rm PML}}} }{\int_{\mathcal{S}_{\rm dipole}}\mathbf{n}_{\mathbf{r}}\cdot {\bf S}_{\rm dipole,total}(\mathbf{r},\omega){\rm d}{A_{\mathbf{r}}}},\label{betaradfull}\\
    \beta_{\rm num}^{\rm nrad}(\mathbf{r}_{0},\omega)=1-\beta_{\rm num}^{\rm rad}(\mathbf{r}_{0},\omega),\label{betanradfull}
\end{gather}
where the surface $\mathcal{S}_{\rm PML}$ is the interface 
just before the PML, the vector ${\bf S}_{\rm PML,total}(\mathbf{r}_{\rm PML},\omega)$ is the Poynting vector and $\mathbf{n}_{\mathbf{r}_{\rm PML}}$ is the normal vector at this interface.
Note that the classical beta factors are frequency dependent
in general, but (in terms of 
the mode of interest) are most important to define near ${\omega_{\rm c}}$.

In a single-QNM case, the form of the commutation relation from Eq.~\eqref{eq: Sfull} (in the limit $\alpha\rightarrow 0$) together Eq.~\eqref{eq: Sin} and Eq.~\eqref{eq: Soutfinomega}/\eqref{eq: Soutfinomega2} simplify as 
\begin{subequations}
\begin{gather}
    S^{\rm nrad}=\frac{2}{\pi\omega_{\rm c}}\int_0^{\infty}{\rm d}\omega|A_{\rm c}(\omega)|^2{\rm Im}[\epsilon_{{\rm PC}}(\omega)]I^{\rm spat}_{\rm in},\label{snrad_full}\\
    S^{\rm rad}=\frac{2}{\pi\omega_{\rm c}}\int_0^{\infty}{\rm d}\omega|A_{\rm c}(\omega)|^2\frac{n_{\rm B}c}{\omega}I^{\rm spat}_{\rm out}, \label{srad_full}\\
    S^{\rm rad\prime}=\frac{2}{\pi\omega_{\rm c}}\int_0^{\infty}{\rm d}\omega|A_{\rm c}(\omega)|^2\frac{n_{\rm B}c}{\omega}I^{\rm spat\prime}_{\rm out}(\omega), \label{srad_full2}
\end{gather}
\end{subequations}
with 
\begin{subequations}
\begin{gather}
I^{\rm spat}_{\rm in}=\int_V {\rm d}\mathbf{r} |\tilde{\mathbf{f}}_{\rm c}^{\rm r}(\mathbf{r})|^2, \\
I^{\rm spat}_{\rm out}=\int_{\mathcal{S}} {\rm d}A_\mathbf{s}{\rm Im}\left[\left( \mathbf{n}_{\mathbf{s}}\times\left(\boldsymbol{\nabla}\times\tilde{\mathbf{f}}_{\rm c}^{\rm r}(\mathbf{s})\right)\right)\cdot\tilde{\mathbf{f}}_{\rm c}^{\rm r*}(\mathbf{s})\right],\\
I^{\rm spat\prime}_{\rm out}(\omega)=\frac{1}{16\pi^2}\int_{0}^{2\pi}d\varphi\int_0^{\pi}\sin(\vartheta) d\vartheta I_{\rm c}^{\rm sur}(\omega),
\end{gather}
\end{subequations}
where we have chosen the surface representation $I_{\rm c}^{\rm sur}(\omega)$ from Eq.~\eqref{I_sur} with $\mu=\eta={\rm c}$. Performing a pole approximation at $\omega=\omega_{\rm c}$, leads then to 
\begin{subequations}
\begin{gather}
    S^{\rm nrad}\approx Q_c{\rm Im}[\epsilon_{{\rm PC}}(\omega_{\rm c})]I^{\rm spat}_{\rm in},\label{snrad_pole} \\
    S^{\rm rad}\approx\frac{n_{\rm B}c}{2\gamma_{\rm c}}I^{\rm spat}_{\rm out},\label{srad_pole}\\
    S^{\rm rad\prime}\approx\frac{n_{\rm B}c}{2\gamma_{\rm c}}I^{\rm spat\prime}_{\rm out}(\omega_{\rm c})\label{Sradpole2}.
\end{gather}
\end{subequations}

The calculation of the $S$ paramaters and the approximated forms involves a volume integral over the absorptive resonator region in $S^{\rm nrad}$ and a surface integral in $S^{\rm rad}$, which are encoded in $I^{\rm spat}_{\rm in/out}$ and $I^{\rm spat\prime}_{\rm out}$. 
See App.~\ref{App: NumPCBeam} for further details.
These quantum-derived  $S$ factors are unitless quantities, and for well isolated single QNMs in metal resonators, we have numerically found that~\cite{PhysRevLett.122.213901,ren_near-field_2020} $S
 =S^{\rm rad}+S^{\rm nrad} {\approx}1$, as we
also find in this work. This is consistent with the derivation of the non-dissipative limit of $S$ in App.~\ref{app: JCmodelBreakdown2}, especially because we are investigating a case with high $Q$ factor.

After obtaining the $S$ parameters and assuming the bad cavity limit, one can obtain the quantum SE rate~\cite{PhysRevLett.122.213901}:
 \begin{equation}
     \Gamma_{\rm quan}=S\frac{2\big|\tilde{g}_{\rm c}\big|^{2}\gamma_{\rm c}}{\Delta_{\rm ce}^{2}+\gamma^{2}_{\rm c}},
     \label{GammaPF}
 \end{equation}
where $\Delta_{\rm ce}=\omega_{\rm c}-\omega_{\rm e}$ is the frequency detuning between the emitter and the single QNM, and $\tilde{g}_{\rm c}=\sqrt{\omega_{\rm c}/(2\epsilon_{0}\hbar)}\mathbf{d}\cdot\tilde{\mathbf{f}}_{\rm c}(\mathbf{r}_{0})$ is the emitter-QNM coupling. Then the quantum Purcell factor is 
\begin{equation}\label{quantumpurcellsingle}
F_{\rm P}^{\rm quan}=\frac{\Gamma_{\rm quan}}{\Gamma_{0}}.  
\end{equation}

In the limit that $S\rightarrow 1$,
Eqs.~\eqref{GammaPF}-\eqref{quantumpurcellsingle} recover the well-known decay rate and Purcell factor from the dissipative Jaynes-Cummings model~\cite{Cirac}, which is also in agreement with the limit of vanishing dissipation of the QNM quantum model, discussed in App.~\ref{app: JCmodelBreakdown2}.
Furthermore, in the quantized QNM theory, the beta factors are defined from 
\begin{align}
 \beta_{\rm quan}^{\rm rad}& =\frac{S^{\rm rad}}{S}, \label{beta_rad_quan}
 \\
 \beta_{\rm quan}^{\rm nrad}&=\frac{S^{\rm nrad}}{S} \label{beta_nrad_quan}. 
\end{align}

\begin{table*}[htb]
\caption {Summary of the calculated classical and quantum QNM parameters for the PC beam cavities: QNM resonance energy $\hbar\tilde{\omega}_{\rm c}$, quality factor $Q_{\rm c}$, 
quantum $S$ parameters (including $S^{\rm nrad}$ from Eq.~\eqref{snrad_pole}, $S^{\rm rad}$ from Eq.~\eqref{srad_pole}, (we also show $S^{\rm rad\prime}$ from Eq.~\eqref{Sradpole2}), and $S=S^{\rm rad}+S^{\rm nrad}$), quantum nonradiative beta factor $\beta_{\rm quan}^{\rm nrad}$ from Eq.~\eqref{beta_nrad_quan} and numerical nonradiative beta factor $\beta_{\rm num}^{\rm nrad}$ (Eq.~\eqref{betanradfull}) from full dipole method for various material losses $\gamma_{\rm p}$ (as well as various ${\rm Im}[\epsilon_{\rm PC}(\omega_{\rm c})]$). When calculating $S^{\rm rad}$, the surface we are choosing is a cuboid, which is 500\,nm and 200\,nm away from the PC beam surface along $x$, $z$ direction and $y$ direction (see App.~\ref{App: NumPCBeam}).
In all cases, we find that $S$ is close to 1, within numerical uncertainty. The quantum nonradiative beta factors $\beta_{\rm quan}^{\rm nrad}$ are also very close to full-dipole classical results $\beta_{\rm num}^{\rm nrad}$.
} \label{S1} 
    \centering
    \begin{tabular}{|c|c|c|c|c|c|c|c|c|}
 \hline
 $\epsilon_{\rm PC}(\gamma_{\rm p})$ & ${\rm Im}[\epsilon_{\rm PC}(\omega_{\rm c})]$ & $\hbar\tilde{\omega}_{c}~[{\rm eV}]$ &  $Q_{\rm c} $ & $S^{\rm nrad}$ & $S^{\rm rad}~(S^{\rm rad\prime})$ & $S$ & $\beta^{\rm nrad}_{\rm quan}$ & $\beta^{\rm nrad}_{\rm num} $ \\
 \hline
 Re[$\epsilon_{\rm PC}(1\gamma_{\rm p0})$]  & $0$ & $(1.5977 - 0.0002985i)$ & $2676$ & $0$ & $1.009$ ($1.001$)  & $1.009$ & $0\%$ & $0\%$  \\
 \hline
 $\epsilon_{\rm PC}(0.1\gamma_{\rm p0})$  & $4.7627\times10^{-4}$ & $(1.5977 - 0.0003662i)$ & $2182$ & $0.193$ & $0.808$ ($0.806$)  & $1.001$ & $19\%$ & $19\%$  \\
 \hline
 $\epsilon_{\rm PC}(0.2\gamma_{\rm p0})$ & $9.5255\times10^{-4}$ & $(1.5977 - 0.0004371i)$ & $1828$ & $0.323$ & $0.677$ ($0.676$)  & $1.000$ & $32\%$ & $32\%$  \\
 \hline
 $\epsilon_{\rm PC}(0.3\gamma_{\rm p0})$ & $1.4288\times10^{-3}$ & $(1.5977 - 0.0005080i)$ & $1573$ & $0.416$ & $0.582$ ($0.583$)  & $0.998$ & $42\%$ & $41\%$ \\
 \hline
 $\epsilon_{\rm PC}(0.5\gamma_{\rm p0})$ & $2.3814\times10^{-3}$ & $(1.5977 - 0.0006499i)$ & $1229$ & $0.539$ & $0.454$ ($0.458$)  & $0.994$ & $54\%$ & $54\%$  \\
 \hline
 $\epsilon_{\rm PC}(1\gamma_{\rm p0})$ & $4.7627\times10^{-3}$ & $(1.5977 - 0.0010044i)$ & $795$ & $0.687$ & $0.292$ ($0.302$)  & $0.979$ & $70\%$ & $69\%$ \\
  \hline
    \end{tabular}
\end{table*}

As a specific example example
for $\gamma_{\rm p}=0.1\gamma_{\rm p0}$,
Fig.~\ref{pc_sche} (b) shows 
 the excellent agreement between the quantum Purcell factor $F_{\rm P}^{\rm quan}$ (Eq.~\eqref{quantumpurcellsingle}) and the numerical Purcell factor $F_{\rm P}^{\rm num}$ (Eq.~\eqref{Purcellfulldipole}) from the full dipole method.
Note that here we show the quantum Purcell factors using $S^{\rm rad}$ obtained from Eq.~\eqref{srad_pole}, and emphasize again, that the results using Eq.~\eqref{Sradpole2} give the same value within the numerical precision.
The corresponding 
QNM fields for $\gamma_{\rm p}=0.1\gamma_{\rm p0}$ in the resonator region are shown in Fig.~\ref{pc_sche} (c) and (d).

 Table~\ref{S1} 
gives a complete summary of the
various $S$ factors and beta factors
as a function of loss including the lossless
limit.
Within the numerical precision, we also find that the total $S=1\pm 0.03$
for the PC beams (with and without material loss), although there is a very slight trend from $S=0.979$ to $S=1.009$ from the case $\gamma_{\rm p}=\gamma_{\rm p0}$ to $\gamma_{\rm p}=0$. Since for all cases $Q\gg 1$, it is clear from the last subsection that $S$ remains close or equal to 1 for the inspected absorption regimes.
Interestingly, as mentioned above,
we also find the same trend for
low loss metal resonators with a Drude model, where 
 $Q\approx 10-20$ \cite{ren_near-field_2020}.
However, $S^{\rm rad}$ and $S^{\rm nrad}$ change drastically in the inspected absorption range towards the limit $\gamma_{\rm p}\rightarrow 0$, as can also be seen in Table~\ref{S1}.

\section{Discussion of our key relations and findings\label{Sec: keyequations}}
In this section, we discuss key relations obtained from the Green function quantization in combination with the introduction of permittivity sequences. In the first step, it was shown that the integral relation 
\begin{align}
\lim_{\alpha\rightarrow 0}\lim_{\lambda\rightarrow \infty}\int_{V(\lambda)}{\rm d}\mathbf{s}~& \epsilon_I^{(\alpha)}(\mathbf{s})\mathbf{G}_{\alpha}(\mathbf{r}, \mathbf{s}) \cdot \bar{\mathbf{G}}_{\alpha}(\mathbf{s}, \mathbf{r}')\nonumber\\
&={\rm Im}\left[\mathbf{G}(\mathbf{r},\mathbf{r}')\right], \label{eq: FDRel1KeyRes}
\end{align}
holds. This is a {\it corrected} version of the results (cf. Eq.~\eqref{eq: GFeps_relation}) from Refs.~\onlinecite{Dung,grunwel}, that holds not only  for absorptive bulk media, but also for, e.g., vacuum case or nanostructures in a lossless environment. In a second step, we have derived a different variant of the relation in Eq.~\eqref{eq: FDRel1KeyRes} for finite nanostructures in 
a lossless background medium:
\begin{align}
\lim_{\alpha\rightarrow 0}&\lim_{\lambda\rightarrow \infty}\int_{V(\lambda)}{\rm d}\mathbf{s}~ \epsilon_I^{(\alpha)}(\mathbf{s})\mathbf{G}_{\alpha}(\mathbf{r}, \mathbf{s}) \cdot \bar{\mathbf{G}}_{\alpha}(\mathbf{s}, \mathbf{r}')\nonumber\\
=&\int_{V_{\rm in}}{\rm d}\mathbf{s}~ \epsilon_{I}(\mathbf{s}) \mathbf{G}(\mathbf{r}, \mathbf{s}) \cdot \mathbf{G}^*(\mathbf{s}, \mathbf{r}')\nonumber\\
&+\frac{c^2}{2i\omega^2}\int_{\mathcal{S}}{\rm d}A_\mathbf{s}~ \Big\{\mathbf{C}(\mathbf{s},\mathbf{r},\mathbf{r}')- \mathbf{C}^\dagger(\mathbf{s},\mathbf{r}',\mathbf{r})\Big\}.\label{eq: Malphafin2KeyRes}
\end{align}

This second relation permits to calculate the zero-point vacuum fluctuation and related quantities, such as the Purcell factor of an atom in dissipative environment, with a finite spatial domain. It also distinguishes the two fundamental dissipative processes more clearly: absorptive loss through the imaginary part of the permittivity and radiation loss through the Poynting vector-like term on a surface surrounding the nanostructure.
Interestingly, by substituting Eq.~\eqref{eq: FDRel1KeyRes} into Eq.~\eqref{eq: Malphafin2KeyRes}, a third relation 
\begin{align}
    {\rm Im}(\mathbf{G}(\mathbf{r},&\mathbf{r}'))=\int_{V_{\rm in}}{\rm d}\mathbf{s}~ \epsilon_{I}(\mathbf{s}) \mathbf{G}(\mathbf{r}, \mathbf{s}) \cdot \mathbf{G}^*(\mathbf{s}, \mathbf{r}')\nonumber\\
&+\frac{c^2}{2i\omega^2}\int_{\mathcal{S}}{\rm d}A_\mathbf{s}~ \Big\{\mathbf{C}(\mathbf{s},\mathbf{r},\mathbf{r}')- \mathbf{C}^\dagger(\mathbf{s},\mathbf{r}',\mathbf{r})\Big\},\label{eq: ImGVandS}
\end{align}
is obtained, where it is important to note, that $\mathbf{r},\mathbf{r}'\in V_{\rm in}$ and $\mathcal{S}$ is strictly finite. Indeed, Eq.~\eqref{eq: ImGVandS} can  be  derived by just integrating the Helmholtz equation of the Green function over a finite volume $V_{\rm in}$ with boundary $\mathcal{S}$ (without the necessity of introducing permittivity sequences), rendering the modified approach consistent with the limit of $\alpha\rightarrow 0$. 

A technically interesting case, is the situation where a two-level 
quantum emitter at position $\mathbf{r}_a$ interacts (weakly) with the surrounding photon field. It can be shown in the framework of the Green function quantization, that the spontaneous emission rate of that quantum emitter is proportional to $\lim_{\alpha\rightarrow 0}\lim_{\lambda\rightarrow \infty}\mathbf{M}_{\alpha,\lambda}(\mathbf{r}_a,\mathbf{r}_a)$. Ref.~\onlinecite{dorier2019critical} pinpoints for Eq.~\eqref{eq: FDRel1KeyRes}, that it seems to not recover the limit of vanishing absorption. However, Eq.~\eqref{eq: Malphafin2KeyRes} does indeed recover this. To make this clear again, we set $\epsilon(\mathbf{r},\omega)=1$ for all space points. It follows, that 
\begin{equation}
    \mathbf{G}(\mathbf{r},\mathbf{r}',\omega)=\mathbf{G}_{\rm B }(\mathbf{r},\mathbf{r}',\omega).
\end{equation}
Using Eq.~\eqref{eq: ImGVandS}, this leads to:
\begin{align}
    {\rm Im}(\mathbf{G}_{\rm B}&(\mathbf{r},\mathbf{r}',\omega))\nonumber\\
    &=\frac{c^2}{2i\omega^2}\oint_{\mathcal{S}} {\rm d}A_\mathbf{s}\left\{\mathbf{C}_{\rm B,\alpha}(\mathbf{s},\mathbf{r},\mathbf{r}')- \mathbf{C}_{\rm B,\alpha}^\dagger(\mathbf{s},\mathbf{r}',\mathbf{r})\right\},
\end{align}
which is indeed consistent with the relation in Eq.~\eqref{eq: Malphafin2KeyRes}, and shows that the limit is preserved. 

Furthermore, Eq.~\eqref{eq: Malphafin2KeyRes} and the other variants of this relation, are the key to obtain the radiative part of the commutation relation of QNM operators: 
\begin{align}
    &\tilde{S}^{\rm rad}_{\mu\eta}(\omega)\nonumber\\
    &=\frac{ic^2}{2\omega^2}\oint_{\mathcal{S}} {\rm d}A_\mathbf{s}\left[\mathbf{n}_{\mathbf{s}}\times\left(\boldsymbol{\nabla}\times\tilde{\mathbf{F}}_{\mu}(\mathbf{s})\right)\right]\tilde{\mathbf{F}}_{\eta}^*(\mathbf{s}){-}{\rm h.c.}(\mu\leftrightarrow\eta),\label{eq: SoutfinomegaKeyRes}
\end{align}
which is consistent with Ref.~\onlinecite{PhysRevLett.122.213901}.
Since the surface $\mathcal{S}$ can be chosen arbitrarily (as long as it remains strictly finite and is far away from the resonator), we have found an exact limit of $\tilde{S}^{\rm rad}_{\mu\eta}(\omega)$, which does not depend on the shape of $\mathcal{S}$ and yields a symmetric, positive and bilinear form 
\begin{equation}
    \tilde{S}^{\rm rad}_{\mu\eta}(\omega)= \frac{\sqrt{\epsilon_{\rm B}(\omega)}c}{\omega}\frac{1}{16\pi^2}\int_{0}^{2\pi}d\varphi\int_0^{\pi}\sin(\vartheta) d\vartheta I_{\mu\eta}(\omega).\label{eq: Soutfin3KeyRes}
\end{equation}
In Eq.~\eqref{eq: Soutfin3KeyRes}, only integrals over the resonator region remain, which improves the modal quantization approach.

\section{Conclusions\label{Sec: Conc}}
In summary, we have reformulated a Green function quantization approach by introducing a sequence of permittivities including a fictitious (lossy) Lorentz oscillator with weight parameter $\alpha$. In the limit $\alpha\rightarrow 0$, the sequence converges to the actual permittivity, allowing us to address explicitly the situation of a lossy or lossless resonator in a infinite lossless background medium. It was shown that the approach, in this form, rigorously recovers the connection of the fluctuation and dissipation, even in the case without absorption. Furthermore, we have explicitly derived a form of the zero-point field fluctuations that includes a volume integral over the absorptive region and a surface term, connected to the radiative dissipation into the lossless background medium, which does not depend on the imaginary part of the permittivity. We have then applied the method to a recent QNM quantization scheme and confirmed a contribution associated to the radiative loss. 

In addition, we have studied a practical and non-trivial numerical example of a lossy photonic-crystal cavity, supporting a single QNM (in some frequency regime of interest). The material model
 includes the dispersion and loss
using a rigorous solution to the
full 3D Maxwell equations, thus combining practical classical and 
quantum calculations on an equal footing.
We then demonstrated how the radiative/non-radiative part of the QNM commutation relation is identical with the radiative/non-radiative $\beta$ factor, obtained via the full Maxwell equation solution, within numerical precision. Calculations for various amounts
of material loss and dispersion are presented (using a causual Lorentz model), including the 
important example of a completely lossless material.
These results shed light on recent controversies that have been raised in the literature about Green function quantization to finite size resonators, and on their own further demonstrate the power of using a QNM approach for rigorous quantization with open
cavity modes.

\acknowledgments
We acknowledge support from the Deutsche Forschungsgemeinschaft (DFG) through SFB 951 Project B12 (Project number 182087777)  and the 
Alexander von Humboldt Foundation through a Humboldt Research Award.
We also acknowledge Queen's University, the Canadian Foundation for Innovation, and the Natural Sciences and Engineering Research Council of Canada for financial support, and CMC Microsystems
for the provision of COMSOL Multiphysics.
We thank 
Philip Tr\o st Kristensen,
Kurt Busch, 
Mohsen Kamandar Dezfouli, and George Hanson for useful discussions.
Very fruitful discussions with Andreas Knorr are gratefully acknowledged.

\appendix
\section{Proof of Eq.~\eqref{eq: GFrel2prime} and Eq.~\eqref{GFrel2}\label{App: ProofMalpha}}
In the following, we show the form of Eq.~\eqref{eq: GFrel2prime} and Eq.~\eqref{GFrel2} by starting with the transposed form of the Green function Helmholtz equation,
\begin{align}
\frac{\omega^2}{c^2}\mathbb{1}\delta(\mathbf{s}-\mathbf{r})=&\left[\boldsymbol{\nabla}_{\mathbf{s}}\times\left[\boldsymbol{\nabla}_{\mathbf{s}}\times\mathbf{G}_{\alpha}(\mathbf{s},\mathbf{r})\right]\right]^{\rm T}\nonumber\\
&-\frac{\omega^2}{c^2}\epsilon^{(\alpha)}(\mathbf{s})\mathbf{G}_{\alpha}(\mathbf{r},\mathbf{s}),
\end{align}
where we have used the reciprocity theorem~\cite{chew1995waves}
\begin{equation}
\left[\mathbf{G}_{\alpha}(\mathbf{s},\mathbf{r})\right]^{\rm T }=\mathbf{G}_{\alpha}(\mathbf{r},\mathbf{s}).
\end{equation}
In addition, the complex conjugated form reads
\begin{align}
\frac{\omega^2}{c^2}\mathbb{1}\delta(\mathbf{s}-\mathbf{r})=&\boldsymbol{\nabla}_{\mathbf{s}}\times\left[\boldsymbol{\nabla}_{\mathbf{s}}\times\bar{\mathbf{G}}_{\alpha}(\mathbf{s},\mathbf{r})\right]\nonumber\\
&-\frac{\omega^2}{c^2}\bar{\epsilon}^{(\alpha)}(\mathbf{s},\omega)\bar{\mathbf{G}}_{\alpha}(\mathbf{s},\mathbf{r}) .
\end{align} 

Next, we look at
\begin{align}
&\mathbf{M}_{\alpha,\lambda}(\mathbf{r},\mathbf{r}')\nonumber\\
&=\frac{1}{2i}\int_{B(\lambda)}{\rm d}\mathbf{s}~ \Big\{\epsilon^{(\alpha)}(\mathbf{s}) \mathbf{G}_{\alpha}(\mathbf{r}, \mathbf{s}) \cdot \bar{\mathbf{G}}_{\alpha}(\mathbf{s}, \mathbf{r}')\nonumber\\
&\hspace{60pt}-\bar{\epsilon}^{(\alpha)}(\mathbf{s}) \mathbf{G}_{\alpha}(\mathbf{r}, \mathbf{s}) \cdot \bar{\mathbf{G}}_{\alpha}(\mathbf{s}, \mathbf{r}')\Big\},
\end{align}
where $B(\lambda)$ is a compact volume with smooth boundary $\partial B(\lambda)$. Using the Helmholtz equations from above, we can rewrite this as 
\begin{align}
&\frac{2i\omega^2}{c^2}\mathbf{M}_{\alpha,\lambda}(\mathbf{r},\mathbf{r}')\\
=&\int_{B(\lambda)}{\rm d}\mathbf{s}~ \Big\{\left[\boldsymbol{\nabla}_{\mathbf{s}}\times\left[\boldsymbol{\nabla}_{\mathbf{s}}\times\mathbf{G}_{\alpha}(\mathbf{s},\mathbf{r})\right]\right]^{\rm T} \cdot \bar{\mathbf{G}}_{\alpha}(\mathbf{s}, \mathbf{r}')\nonumber\\
&- \left[\mathbf{G}_{\alpha}(\mathbf{s}, \mathbf{r})\right]^{\rm T} \cdot \left[\boldsymbol{\nabla}_{\mathbf{s}}\times\left[\boldsymbol{\nabla}_{\mathbf{s}}\times\bar{\mathbf{G}}_{\alpha}(\mathbf{s},\mathbf{r})\right]\right]\Big\}\nonumber\\
&-\frac{\omega^2}{c^2}\left[\mathbf{I}_\alpha(\mathbf{r},\mathbf{r}')-\mathbf{I}_\alpha^*(\mathbf{r}',\mathbf{r})\right],
\end{align}
where 
\begin{equation}
\mathbf{I}_\alpha(\mathbf{r},\mathbf{r}') =\int_{B(\lambda)} {\rm d}\mathbf{r}\delta(\mathbf{s}-\mathbf{r})\mathbf{G}_{\alpha}^*(\mathbf{s},\mathbf{r}').
\end{equation}

Furthermore, since $B(\lambda)$ is a volume with smooth surface $\partial B(\lambda)$ and compact (at least for finite $\lambda> 0$), we can use the Green second dyadic-dyadic identity~\cite{martin2006multiple}: 
\begin{align}
\int_B &\left\{\left[\boldsymbol{\nabla}\times\left[\boldsymbol{\nabla}\times\mathbf{Q}\right]\right]^{\rm T} \cdot \mathbf{P}- \left[\mathbf{Q}\right]^{\rm T} \cdot \left[\boldsymbol{\nabla}\times\left[\boldsymbol{\nabla}\times\mathbf{P}\right]\right]\right\}\nonumber\\
&=\oint_{\partial B} \left\{\left[\mathbf{n}\times \mathbf{Q}\right]^{\rm T} \cdot \left[\boldsymbol{\nabla}\times \mathbf{P}\right] - \left[\boldsymbol{\nabla}\times \mathbf{Q}\right]^{\rm T} \cdot \left[\mathbf{n}\times \mathbf{P}\right]\right\}, \label{eq: DyadDyadGF}
\end{align}
to arrive at 
\begin{align}
\mathbf{M}_{\alpha}&(\mathbf{r},\mathbf{r}')=\frac{i}{2}\left[\mathbf{I}_\alpha(\mathbf{r},\mathbf{r}')-\mathbf{I}_\alpha^*(\mathbf{r}',\mathbf{r})\right]\nonumber\\
&+\frac{c^2}{2i\omega^2}\lim_{\lambda\rightarrow\infty}\int_{\partial B(\lambda)}{\rm d}A_\mathbf{s}\left\{\mathbf{C}_{\alpha}(\mathbf{s},\mathbf{r},\mathbf{r}'){-}\mathbf{C}_{\alpha}^\dagger(\mathbf{s},\mathbf{r}',\mathbf{r})\right\}.
\end{align}
Taking the limit $\lambda\rightarrow\infty$ in $\mathbf{I}_\alpha(\mathbf{r},\mathbf{r}')$, and applying the Dirac delta function, gives 
\begin{equation}
    \frac{i}{2}\left[\mathbf{I}_\alpha(\mathbf{r},\mathbf{r}')-\mathbf{I}_\alpha^*(\mathbf{r}',\mathbf{r})\right]=\begin{cases}
    {\rm Im}\left[\mathbf{G}_{\alpha}(\mathbf{r},\mathbf{r}')\right]\quad\mathbf{r},\mathbf{r}'\in B\\
    0\quad \mathbf{r},\mathbf{r}'\notin B
    \end{cases},
\end{equation}
where $B=\lim_{\lambda\rightarrow\infty}B(\lambda)$.
Choosing $B(\lambda)=V(\lambda)$ leads to the form in Eq.~\eqref{eq: GFrel2prime}, while  $B(\lambda)=V_{\rm out}(\lambda)$ leads to Eq.~\eqref{GFrel2}.
\section{Proof of Eq.~\eqref{eq: SinfInt}\label{App: GFHom}}

We start with the Dyson equation, which reads
\begin{align}
    \mathbf{G}_{\alpha}(\mathbf{s},\mathbf{r})=&\mathbf{G}_{\rm B,\alpha}(\mathbf{s},\mathbf{r})\nonumber\\
    &+\int_V {\rm d}\mathbf{r}' \Delta\epsilon^{(\alpha)}(\mathbf{r}')\mathbf{G}_{\rm B,\alpha}(\mathbf{s},\mathbf{r}')\cdot\mathbf{G}_{\alpha}(\mathbf{r}',\mathbf{r}).
\end{align}
Next, we note that $\mathbf{G}_{\rm B,\alpha}(\mathbf{r},\mathbf{r}')$ is defined through~\cite{Hecht}
\begin{equation}
    \mathbf{G}_{\rm B,\alpha}(\mathbf{s},\mathbf{r})=k_0^2\left[\mathbb{1}-\frac{\boldsymbol{\nabla}_{\mathbf{s}}\boldsymbol{\nabla}_{\mathbf{s}}}{k_\alpha^2}\right]g_{\rm B,\alpha}(|\mathbf{s}-\mathbf{r}|),
\end{equation}
with 
\begin{equation}
    g_{\rm B,\alpha}(|\mathbf{s}-\mathbf{r}|)=\frac{e^{ik_\alpha |\mathbf{s}-\mathbf{r}|}}{4\pi |\mathbf{s}-\mathbf{r}|}, \label{eq: smallghom}
\end{equation}
 where $k_\alpha=\sqrt{\epsilon^{(\alpha)}_{\rm B}}\omega/c$.
Applying the differential operators on $g_{\rm B,\alpha}(|\mathbf{s}-\mathbf{r}|)$ leads to~\cite{Hecht, chew1995waves}
\begin{align}
    \mathbf{G}_{\rm B,\alpha}(\mathbf{s},\mathbf{r}) = k_0^2\Bigg[&\left(1+\frac{ik_\alpha R -1}{k_\alpha^2 R^2}\right)\mathbb{1}\nonumber\\
    &+\frac{3-3ik_\alpha R -k_\alpha^2 R^2}{k_\alpha^2 R^4}\mathbf{R}\mathbf{R}\Bigg]g_{\rm B,\alpha}(R), \label{eq: GFBfull}
\end{align}
where $R$ is the absolute value of the distance vector $\mathbf{R}=\mathbf{s}-\mathbf{r}$.

Analogously, $\boldsymbol{\nabla}_{\mathbf{s}}\times \mathbf{G}_{\rm B,\alpha}(\mathbf{s},\mathbf{r})$ is
\begin{equation}
    \boldsymbol{\nabla}_{\mathbf{s}}\times \mathbf{G}_{\rm B,\alpha}(\mathbf{s},\mathbf{r})=k_0^2\frac{\mathbf{R}\times\mathbb{1}}{R}\left(i-\frac{1}{k_\alpha R}\right)g_{\rm B,\alpha}(R), 
\end{equation}
and $\mathbf{R}\times\mathbb{1}$ is a dyad, given as the cross product of $\mathbf{R}$ with every column of the unit matrix $\mathbb{1}$. In general the prefactor of $\mathbf{G}_{\rm B,\alpha}(\mathbf{s},\mathbf{r})$ can be split into three parts scaling with $R^{-3}$, $R^{-2}$ and $R^{-1}$, corresponding to near-field, intermediate field and far-field contributions, respectively. 
Since $\lambda\rightarrow \infty$ is equivalent to $|\mathbf{s}|\rightarrow\infty$ in Eq.~\eqref{eq: GFrel2prime}, we only take the far field contribution into account and expand
\begin{equation}
|\mathbf{s}-\mathbf{r}|=|\mathbf{s}|\sqrt{1-2\frac{\hat{\mathbf{s}}\cdot\mathbf{r}}{|\mathbf{s}|}+\frac{|\mathbf{r}|^2}{|\mathbf{s}|^2}}\rightarrow |\mathbf{s}| - \hat{\mathbf{s}}\cdot\mathbf{r} + \frac{|\mathbf{r}|^2}{2|\mathbf{s}|},
\end{equation}
where $\hat{\mathbf{s}}=\mathbf{s}/|\mathbf{s}|$ is a unit vector in the direction of $\mathbf{s}$. In the denominator of $g^{\rm B}_\alpha(R)$ (Eq.~\eqref{eq: smallghom}), only the first term in the Taylor expansion $|\mathbf{s}|$ plays a significant role. However, in the fast oscillating exponential function, also the second term $- \hat{\mathbf{s}}\cdot\mathbf{r}$ must be taken into account. The third term $|\mathbf{r}|^2/(2|\mathbf{s}|)$ can be neglected in both contributions, as long as 
\begin{equation}
\frac{k|\mathbf{r}|^2}{2|\mathbf{s}|}\ll 1 
\end{equation}
holds for any finite $\mathbf{r}$, which is the case here. This leads to
\begin{equation}
g^{\rm B}_\alpha(\mathbf{R},\mathbf{r},\omega) \rightarrow \frac{1}{4\pi |\mathbf{s}|}e^{ik_\alpha|\mathbf{s}|}e^{-ik_\alpha(\hat{\mathbf{s}}\cdot\mathbf{r})}.\label{approxsmallG}
\end{equation}

Furthermore, we note that $\hat{\mathbf{R}}=\mathbf{s}/R-\mathbf{r}/R$, in the limit of $|\mathbf{s}|\rightarrow\infty$, yields the limit $\hat{\mathbf{R}}\rightarrow \hat{\mathbf{s}}$. The total far field contribution can therefore be written as
\begin{align}
    \mathbf{G}_{\rm B,\alpha}(\mathbf{s},\mathbf{r})&\rightarrow k_0^2\left[\mathbb{1}-\hat{\mathbf{s}}\hat{\mathbf{s}}\right]\frac{e^{ik_\alpha |\mathbf{s}|}}{4\pi |\mathbf{s}|}e^{-ik_\alpha \hat{\mathbf{s}}\cdot\mathbf{r}}\nonumber\\
    &\equiv \frac{e^{ik_\alpha |\mathbf{s}|}}{4\pi |\mathbf{s}|}\mathbf{K}_\alpha^{(1)}(\hat{\mathbf{s}},\mathbf{r}).\label{eq: K1func}
\end{align}
Similarly, we derive for the curl of  $\mathbf{G}_{\rm B,\alpha}(\mathbf{s},\mathbf{r})$ 
\begin{align}
    \boldsymbol{\nabla}_{\mathbf{s}}\times \mathbf{G}_{\rm B,\alpha}(\mathbf{s},\mathbf{r})&\rightarrow ik_0^2\hat{\mathbf{s}}\times\mathbb{1}\frac{e^{ik_\alpha |\mathbf{s}|}}{4\pi |\mathbf{s}|}e^{-ik_\alpha \hat{\mathbf{s}}\cdot\mathbf{r}}\nonumber\\
    &\equiv\frac{e^{ik_\alpha |\mathbf{s}|}}{4\pi |\mathbf{s}|}\mathbf{K}_\alpha^{(2)}(\hat{\mathbf{s}},\mathbf{r}).\label{eq: K2func}
\end{align}
Therefore, we can rewrite the Dyson equation for  $|\mathbf{s}|\rightarrow\infty$ as
\begin{align}
    \mathbf{G}_{\alpha}(\mathbf{s},\mathbf{r})=&\frac{e^{ik_\alpha |\mathbf{s}|}}{4\pi |\mathbf{s}|}\mathbf{L}_{\alpha}^{(1)}(\hat{\mathbf{s}},\mathbf{r}), \label{eq: GfarDys}
\end{align}
and after the applying the curl as
\begin{align}
   \boldsymbol{\nabla}_{\mathbf{s}}\times \mathbf{G}_{\alpha}(\mathbf{s},\mathbf{r})=&\frac{e^{ik_\alpha |\mathbf{s}|}}{4\pi |\mathbf{s}|}\mathbf{L}_{\alpha}^{(2)}(\hat{\mathbf{s}},\mathbf{r}),\label{eq: RotGfarDys}
\end{align}
with
\begin{align}
    \mathbf{L}_{\alpha}^{(i)}(\hat{\mathbf{s}},\mathbf{r})=&\mathbf{K}_{\alpha}^{(i)}(\hat{\mathbf{s}},\mathbf{r})\nonumber\\
    &+\int_V {\rm d}\mathbf{r}' \Delta\epsilon^{(\alpha)}(\mathbf{r}')\mathbf{K}_{\alpha}^{(i)}(\hat{\mathbf{s}},\mathbf{r}')\cdot\mathbf{G}_{\alpha}(\mathbf{r}',\mathbf{r}),\label{eq: Lfunctions}
\end{align}
for $i=1,2$. Next, we insert Eq.~\eqref{eq: GfarDys} and \eqref{eq: RotGfarDys} into Eq.~\eqref{eq: C_Def}. and the resulting relation then into Eq.~\eqref{eq: GFrel2prime}. Choosing spherical coordinates with respect to $\mathbf{s}$ finally leads to Eq.~\eqref{eq: SinfInt}.

\section{Non-dissipative limit of $S_{\mu\eta}$\label{app: JCmodelBreakdown2}}

In the following, we look at the limit of $S_{\mu\eta}$ from Eq.~\eqref{eq: Sfull} without damping in the system. We analyze the non-radiative and radiative contribution of $S_{\mu\eta}$ separately; the part describing ohmic loss (cf. Eq.~\eqref{eq: Sin}), 
\begin{equation}
    S_{\mu\eta}^{\rm nrad}=\frac{2}{\pi\sqrt{\omega_\mu\omega_\eta}}\int_0^\infty{\rm d}\omega A_\mu(\omega)A_{\eta}^*(\omega)\tilde{S}_{\mu\eta}^{\rm nrad}(\omega),
\end{equation}
 immediately vanishes in a non-absorptive cavity, since the supporting domain of $\epsilon_I(\mathbf{r})$ vanishes; however, the limit of the radiative part (cf.~\eqref{eq: Soutfinomega2}),
\begin{equation}
    S_{\mu\eta}^{\rm rad}=\frac{2}{\pi\sqrt{\omega_\mu\omega_\eta}}\int_0^\infty{\rm d}\omega A_\mu(\omega)A_{\eta}^*(\omega)\tilde{S}_{\mu\eta}^{\rm rad}(\omega),
\end{equation}
 is more subtle; we first rewrite this part
as
\begin{equation}
    S_{\mu\eta}^{\rm rad} = \int_0^\infty{\rm d}\omega\frac{n_{\rm B}c\tilde{L}_{\mu\eta}(\omega)}{i(\tilde{\omega}_\mu-\tilde{\omega}_\eta^*)} \int_{\mathcal{S}_{\rm far}} {\rm d}\mathbf{s}\tilde{\mathbf{F}}_{\mu}(\mathbf{s},\omega)\cdot\tilde{\mathbf{F}}_{\eta}^*(\mathbf{s},\omega),\label{eq: SradDiff}
\end{equation}
with the generalized Lorentzian 
\begin{equation}
    \tilde{L}_{\mu\eta}(\omega)=\frac{\omega}{\sqrt{\omega_\mu\omega_\eta}}\frac{1}{2\pi}\frac{i(\tilde{\omega}_\mu-\tilde{\omega}_\eta^*)}{(\omega-\tilde{\omega}_\mu)(\omega-\tilde{\omega}_\eta^*)}.
\end{equation}

We now inspect the case $\mu\neq\eta$. Here the denominator $i(\tilde{\omega}_\mu-\tilde{\omega}_\eta^*)$ with $\tilde{\omega}_\mu=\omega_\mu-i\gamma_\mu$ remains finite when taking the limit $\gamma_\mu\rightarrow 0$, while the surface integral in the numerator vanishes, since the surface sources in the definition of $\tilde{\mathbf{F}}_{\mu}(\mathbf{s},\omega)$ from Eq.~~\eqref{eq: FregFEP} vanish, and therefore $S_{\mu\eta}^{\rm rad}\rightarrow 0$ for $\mu\neq\eta$. What remains to show is that $S_{\mu\mu}\rightarrow 1$ for $\gamma_\mu\rightarrow 0$. The diagonal elements read explicitly as
\begin{equation}
    S_{\mu\mu} = \frac{n_{\rm B}c}{2\gamma_\mu}\int_0^\infty{\rm d}\omega\tilde{L}_{\mu\mu}(\omega)\int_{\mathcal{S}_{\rm far}} {\rm d}\mathbf{s}|\tilde{\mathbf{F}}_{\mu}(\mathbf{s},\omega)|^2.
\end{equation}
Since we inspect the limit $\gamma_\mu\rightarrow 0$, we can use the property of the regularized field $\tilde{\mathbf{F}}_{\mu}(\mathbf{s},\tilde{\omega}_\mu)= \tilde{\mathbf{f}}_{\mu}(\mathbf{s})$ and since $\tilde{\omega}_\mu\rightarrow \omega_\mu$, and the frequency integral in the above equation is dominated by the Lorentzian at $\omega=\omega_\mu$.
Thus we can approximate $S_{\mu\mu}$ as
\begin{equation}
    S_{\mu\mu} \approx \frac{n_{\rm B}c}{2\gamma_\mu}\int_{\mathcal{S}_{\rm far}} {\rm d}\mathbf{s}|\tilde{\mathbf{f}}_{\mu}(\mathbf{s})|^2.
\end{equation}

Next, we look into the definition of the bilinear form of QNMs for non-absorptive media with spatial-homogeneous background~\cite{KristensenHughes},
\begin{equation}
    \langle \tilde{\mathbf{f}}_{\mu},\tilde{\mathbf{f}}_{\eta} \rangle = \int_{V}\epsilon(\mathbf{r})\tilde{\mathbf{f}}_{\mu}(\mathbf{r})\cdot\tilde{\mathbf{f}}_{\eta}(\mathbf{r}) + \frac{in_{\rm B}c}{\tilde{\omega}_\mu+\tilde{\omega}_\eta} \int_{\mathcal{S}_{\rm far}}{\rm d}\mathbf{s}\tilde{\mathbf{f}}_{\mu}(\mathbf{s})\cdot \tilde{\mathbf{f}}_{\eta}(\mathbf{s}).\label{eq: BilinFormQNM}
\end{equation}
Using the properties $\tilde{\omega}_{-\mu}=-\tilde{\omega}_\mu^*$ and $\tilde{\mathbf{f}}_{-\mu}=\tilde{\mathbf{f}}_{\mu}^*$, and the orthogonalization $ \langle \tilde{\mathbf{f}}_{\mu},\tilde{\mathbf{f}}_{\eta} \rangle=\delta_{\mu\eta}$, we arrive at (with $\eta=-\mu$):
\begin{equation}
    0=\int_{V}\epsilon(\mathbf{r})|\tilde{\mathbf{f}}_{\mu}(\mathbf{r})|^2 -\frac{n_{\rm B}c}{2\gamma_\mu} \int_{\mathcal{S}_{\rm far}}{\rm d}\mathbf{s}|\tilde{\mathbf{f}}_{\mu}(\mathbf{s})|^2.
\end{equation}

Therefore, $S_{\mu\mu}^{\rm rad}=S_{\mu\mu}$ can be rewritten as
\begin{equation}
    S_{\mu\mu} =  \int_{V}\epsilon(\mathbf{r})|\tilde{\mathbf{f}}_{\mu}(\mathbf{r})|^2. \label{eq: NormNormalM}
\end{equation}
In the limit $\gamma_\mu\rightarrow 0$, $\tilde{\mathbf{f}}_{\mu}$ becomes a solution to the Helmholtz equation with closed boundary conditions. Furthermore, the second term on the rhs of the bilinear form, Eq.~\eqref{eq: BilinFormQNM} vanishes in this limit and $\langle \tilde{\mathbf{f}}_{\mu},\tilde{\mathbf{f}}_{\mu} \rangle=1$, and this is also equal to $S_{\mu\mu}$ in the form of Eq.~\eqref{eq: NormNormalM}, which proves that $S_{\mu\mu}\rightarrow 1$ for all $\mu$ in the limit $\gamma_\mu\rightarrow 0$. We note, that using a completely different quantization procedure for 1-dimensional dielectric cavities, it was also shown, that the photon commutation relation approaches 1 in the limit of vanishing cavity leakage.~\cite{2ndquanho}.

Consequently, we have explicitly shown that the QNM quantization model gives the correct limit for the non-dissipative case, namely the well known quantization for closed resonators.

 \section{Consistency check of the QNM Green function expansion\label{app: GFQNMQuant}}

We start with the fundamental commutation relation, $\lim_{\alpha\rightarrow 0}[E_{i,\alpha}(\mathbf{r}),B_{j,\alpha}(\mathbf{r}')]_-$, that can be calculated upon using Eq.~\eqref{eq: FDRel1} and the symmetry properties of the Green function~\cite{scheel1998qed}:
\begin{align}
    \lim_{\alpha\rightarrow 0}[&E_{i,\alpha}(\mathbf{r}),B_{j,\alpha}(\mathbf{r}')]_- \nonumber\\
    &= \frac{\hbar}{\pi\epsilon_0}\sum_m\epsilon_{imk}\partial_m^{\mathbf{r}}\int_{-\infty}^\infty {\rm d}\omega~\frac{1}{\omega}G_{kj}(\mathbf{r},\mathbf{r}',\omega).
\end{align}
Here, $\mathbf{G}(\mathbf{r},\mathbf{r}',\omega)$ is in general a linear combination of a transverse part (or quasi-transverse) $\mathbf{G}^{\perp}(\mathbf{r},\mathbf{r}',\omega)$ and a longitudinal part $\mathbf{G}^{\parallel}(\mathbf{r},\mathbf{r}',\omega)$. 

We first consider positions $\mathbf{r}$ and $\mathbf{r}'$ in the resonator region. As discussed already, the transverse part of the total Green function for these positions is well approximated by the modal QNM Green function from Eq.~\eqref{eq: GFQNM}.
Hence the longitudinal part of the total Green function is simply the longitudinal part of the non-scattering part, i.e., the background Green function $\mathbf{G}_{\rm B}(\mathbf{r},\mathbf{r}',\omega)$:
\begin{align}
    \mathbf{G}^{\parallel}_{\rm B}(\mathbf{r},\mathbf{r}',\omega) =  -\frac{1}{4\pi\epsilon_{\rm B}}\left(\frac{4\pi}{3}\delta(\mathbf{R})\mathbb{1}+\frac{1}{R^3}\left[\mathbb{1}-\frac{3\mathbf{R}\mathbf{R}}{R^2} \right]\right), 
\end{align}
where $\mathbf{R}=\mathbf{r}-\mathbf{r}'$ and $R=|\mathbf{R}|$.
For positions $\mathbf{r},\mathbf{r}'$ in the scattering region, we therefore use the QNM Green function from Eq.~\eqref{eq: GFQNM} to get, as a first contribution 
\begin{align}
\int_{-\infty}^\infty {\rm d}\omega~\frac{1}{\omega}G^{\perp}_{kj}(\mathbf{r},\mathbf{r}',\omega)&= \frac{1}{2}\sum_\mu L_\mu \tilde{f}_{\mu,k}(\mathbf{r})\tilde{f}_{\mu,j}(\mathbf{r}'), 
\end{align}
where 
\begin{align}
  L_\mu&=\int_{-\infty}^\infty {\rm d}\omega \frac{1}{\tilde{\omega}_\mu-\omega}\\
  &= \int_{-\infty}^\infty {\rm d}\omega \frac{\omega_\mu-\omega}{|\tilde{\omega}_\mu-\omega|^2} +i\int_{-\infty}^\infty {\rm d}\omega\frac{\gamma_\mu}{|\tilde{\omega}_\mu-\omega|^2} .\label{eq: FreqIntCommute}
\end{align}

Performing the substitution $\Omega=\omega_\mu-\omega$, we see that the first contribution of the RHS of Eq.~\eqref{eq: FreqIntCommute} vanishes, since the integrated function is antisymmetric. The remaining contribution is a Lorentz function (unnormalized), so that $ L_\mu=i\pi$. It follows that
\begin{align}
 \int_{-\infty}^\infty {\rm d}\omega~\frac{1}{\omega}G^{\perp}_{kj}(\mathbf{r},\mathbf{r}',\omega)&= i\pi\frac{1}{2}\sum_\mu  \tilde{f}_{\mu,k}(\mathbf{r})\tilde{f}_{\mu,j}(\mathbf{r}')\\
&=i\pi\delta_{kj}\delta(\mathbf{r}-\mathbf{r}'),
\end{align}
 where in the last line we used the completeness relation of QNMs~\cite{Leung3,MDR1} under the condition $\epsilon(\mathbf{r},\Omega)\rightarrow 1$ for $|\Omega|\rightarrow \infty$.
Next, we derive the longitudinal contribution. 
We note that, for a lossless background medium, $\mathbf{G}^{\parallel}(\mathbf{r},\mathbf{r}',\omega)/\omega$ has only a simple pole at $\omega=0$  and according to Ref.~\onlinecite{Dung}, we evaluate the frequency integral as a principal value:
\begin{align}
\int_{-\infty}^\infty {\rm d}\omega~\frac{1}{\omega}G^{\parallel}_{kj}(\mathbf{r},\mathbf{r}',\omega)\rightarrow G^{\parallel}_{{\rm B},kj}(\mathbf{r},\mathbf{r}')\mathcal{P}\int_{-\infty}^\infty {\rm d}\omega \frac{1}{\omega} =0.
\end{align}
Therefore, the total commutation relation for $\mathbf{r},\mathbf{r}'$ in the resonator volume reads 
\begin{equation}
    \lim_{\alpha\rightarrow 0}[E_{i,\alpha}(\mathbf{r}),B_{j,\alpha}(\mathbf{r}')]_- = -\frac{i\hbar}{\epsilon_0}\sum_m\epsilon_{ijm}\partial_m^{\mathbf{r}}\delta(\mathbf{r}-\mathbf{r}') ,\label{eq: CommuteEBinner}
\end{equation}
in full agreement with the fundamental commutation relation from free space QED.

If $\mathbf{r}$ and/or $\mathbf{r}'$ are not in the resonator region, then we use again the Dyson equation 
\begin{equation}
    \mathbf{G}(\mathbf{r},\mathbf{r}')= \mathbf{G}_{\rm B}(\mathbf{r},\mathbf{r}')+\int_{V_{\rm in}}{\rm d}\mathbf{s}\Delta\epsilon(\mathbf{s})\mathbf{G}_{\rm B}(\mathbf{r},\mathbf{s})\mathbf{G}(\mathbf{s},\mathbf{r}') ,\label{eq: Dyson}
\end{equation}
to obtain the correct total Green function
for this spatial situation. It was shown in Ref.~\onlinecite{scheel1998qed}, that only the isolated background part $\mathbf{G}_{\rm B}(\mathbf{r},\mathbf{r}')$ then contributes to the commutation relation, if the propagator in the integral kernel ($\mathbf{G}(\mathbf{r},\mathbf{r}')$ in the second term of Eq.~\eqref{eq: Dyson}) satisfies all analytical properties of a Green function, which is the case for the QNM Green function. Hence, in this case, we obtain
\begin{align}
   \lim_{\alpha\rightarrow 0}[&E_{i,\alpha}(\mathbf{r}),B_{j,\alpha}(\mathbf{r}')]_- \nonumber\\
   &= \frac{\hbar}{\pi\epsilon_0}\sum_m\epsilon_{imk}\partial_m^{\mathbf{r}}\int_{-\infty}^\infty {\rm d}\omega~\frac{1}{\omega}G_{{\rm B,}kj}(\mathbf{r},\mathbf{r}',\omega),
\end{align}
which can be shown to be identical to Eq.~\eqref{eq: CommuteEBinner}  using the analytical form of $G_{{\rm B,}kj}(\mathbf{r},\mathbf{r}',\omega)$. This completes the consistency check of the QNM expansion and the specific form of the coefficient $A_\mu(\omega)$ in connection with the Green function quantization approach.

\section{Numerical calculations of the QNM and $S$ values for the photonic crystal beam cavity\label{App: NumPCBeam}}

In this appendix, we present further details on
the numerical calculations of the QNMs,
quantization $S$ factors, and Purcell factors. 
As shown in Fig.~\ref{pc_sche}(a), we consider a 
practical example of a photonic crystal (PC) beam
 similar to that in Refs.~\onlinecite{KamandarDezfouli2017,ren_near-field_2020}.
Compared with the PC beam used in Ref.~\onlinecite{ren_near-field_2020}, the number of the air holes in the mirror region is decreased from $7$ to $3$,
although there are still $7$ holes in the taper region.
The width and height of the PC beam are $W_{\rm PC}=376$ nm and $h_{\rm PC}=200$ nm, respectively. The length of the PC beam was set as $6.052~\mu$m to 
more easily simulate a finite size for the scattering geometry. 
The nanobeam includes a mirror and a taper region, where the taper section was made of $7$ air holes with radius increased from $68$ nm to $86$ nm and their spacing increased from $264$ nm to $299$ nm, in a linear fashion. At the ends of the taper section, $3$ air holes with fixed radius $86$ nm and fixed spacing $306$ nm were used in the mirror region. The length of the cavity region in between the two smallest holes, i.e., in the very middle of the structure, was set as $126$ nm.

We performed the QNM simulations in a commercial COMSOL software~\cite{comsol}, where a computational cylindrical domain of approximately $156~\mu$m$^3$ (including perfectly matched layers (PMLs)) was used with a maximum mesh size of 40 nm and 111 nm on the PC beam and free space, respectively. The $z$-polarized dipole is surrounded by a small sphere with a radius of $1$ nm, where the maximum mesh size is $0.1$ nm. We also used $10$ PMLs to minimize boundary reflections.

Numerically, we obtained the scattered Green function from the scattered electric field of a point dipole source with dipole moment $\bf d$ at location ${\bf r}_0$, by solving the full Maxwell equations in real frequency space.
Assuming $\omega$ is close to $\omega_{\rm c}$, the scattered field from the dipole
is obtained from
\begin{equation}
\mathbf{E}^{\rm s}(\mathbf{r},\omega)=\frac{1}{\epsilon_{0}}\mathbf{G}(\mathbf{r},\mathbf{r}_{0},\omega)\cdot \mathbf{d}.
\end{equation}
If only a single QNM is dominated, one can expand the scattered Green function via (same as Eq.~\eqref{eq:GFwithSUM})
\begin{equation}\label{G_scatter}
\mathbf{G}\left(\mathbf{r},\mathbf{r}_{0},\omega\right)=A_{\rm c}\left(\omega\right)\,\tilde{\mathbf{f}}^{\rm r}_{\rm c}\left(\mathbf{r}\right)\tilde{\mathbf{f}}_{\rm c}^{\rm r}\left(\mathbf{r}_{0}\right).
\end{equation}
Multiplying the left and right sides of $\mathbf{G}\left(\mathbf{r},\mathbf{r}_{0},\omega\right)$ by $\mathbf{d}$, same as the right side of the Eq.~\eqref{G_scatter}, and using 
$\mathbf{r}=\mathbf{r}_{0}$, we then 
obtain the complex rQNM field value at the dipole location, from an inverse Green function expansion over one mode:
\begin{equation}
\tilde{\mathbf{f}}^{\rm r}_{\rm c}\left(\mathbf{r}_{0}\right)\cdot\mathbf{d}=\sqrt{\frac{\mathbf{d}\cdot\mathbf{G}\left(\mathbf{r}_{0},\mathbf{r}_{0},\omega\right)\cdot\mathbf{d}}{A_{\rm c}\left(\omega\right)}}.\label{eq:fnut}
\end{equation}

Substituting this back to Eq.~\eqref{G_scatter}, we easily determine the rQNM, properly normalized, from
\begin{equation}
\tilde{\mathbf{f}}^{\rm r}_{\rm c}\left(\mathbf{r}\right)=\frac{\mathbf{G}\left(\mathbf{r},\mathbf{r}_{0},\omega\right)\cdot\mathbf{d}}{\sqrt{A_{\rm c}\left(\omega\right)\left[\mathbf{d}\cdot\mathbf{G}\left(\mathbf{r}_{0},\mathbf{r}_{0},\omega\right)\cdot\mathbf{d}\right]}},\label{eq:QNM}
\end{equation}
at all spatial positions in the simulation volume.
As shown in Ref.~\onlinecite{RegQNMs}, the real part of the Green function is problematic for obtaining transverse system modes;
however, one can  
use an efficient solution to this problem by using only the (well-behaved) imaginary part of the Green function at two different real frequency points ($\omega_{1}$ and $\omega_{2}$, such as locating at either side of the resonance frequency $\omega_{\rm c}$) to reconstruct the normalized transverse field. The details can be seen from Eqs. (10)-(15) in Ref.~\onlinecite{RegQNMs} (which use the finite-difference time-domain technique, which could also be used here). Here we follow the same procedure to obtain the regularized fields $\tilde{\mathbf{f}}^{\rm r}_{\rm c}\left(\mathbf{r}\right)$.

\begin{figure*}
  \includegraphics[width=0.65\columnwidth]{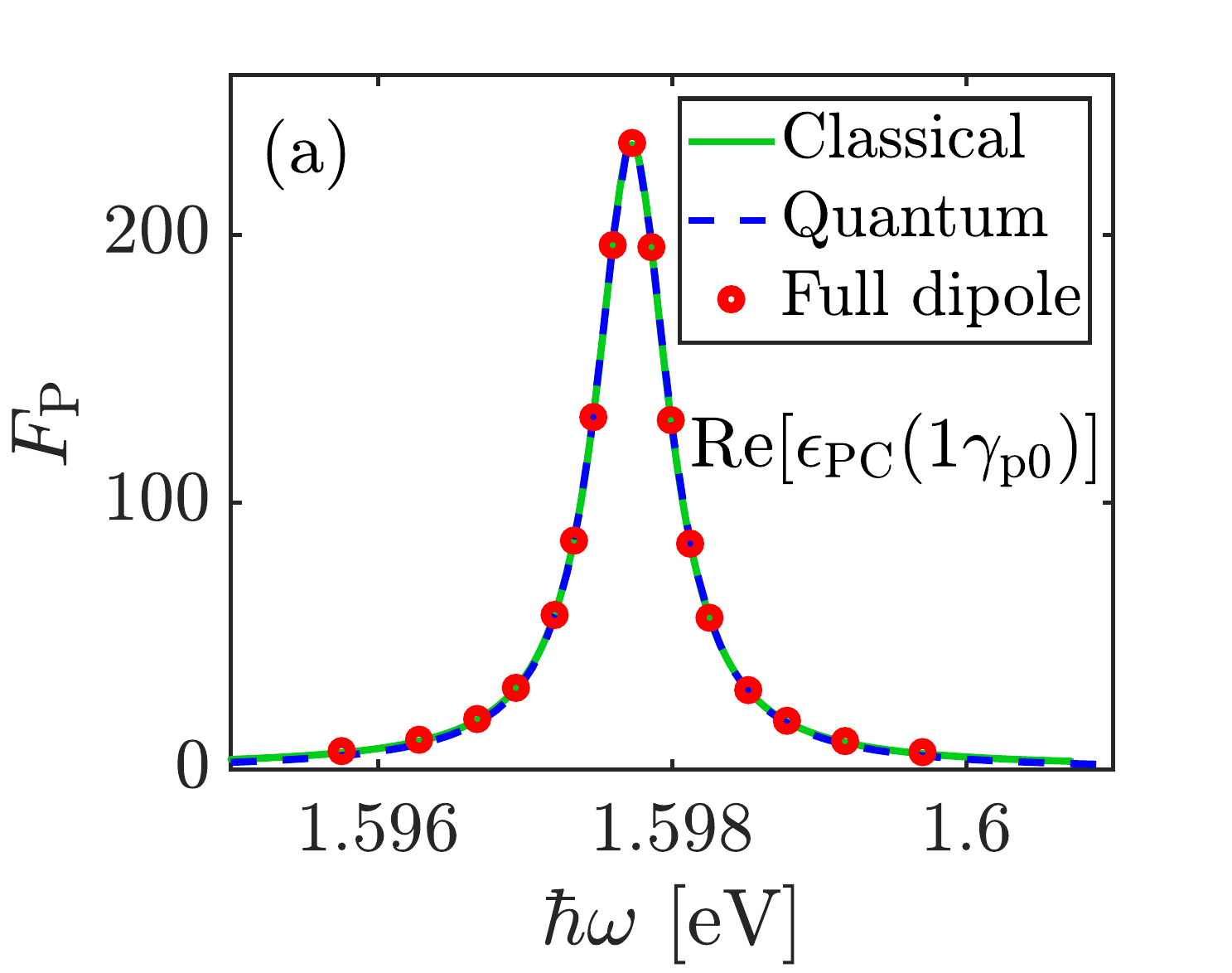}
   \includegraphics[width=0.65\columnwidth]{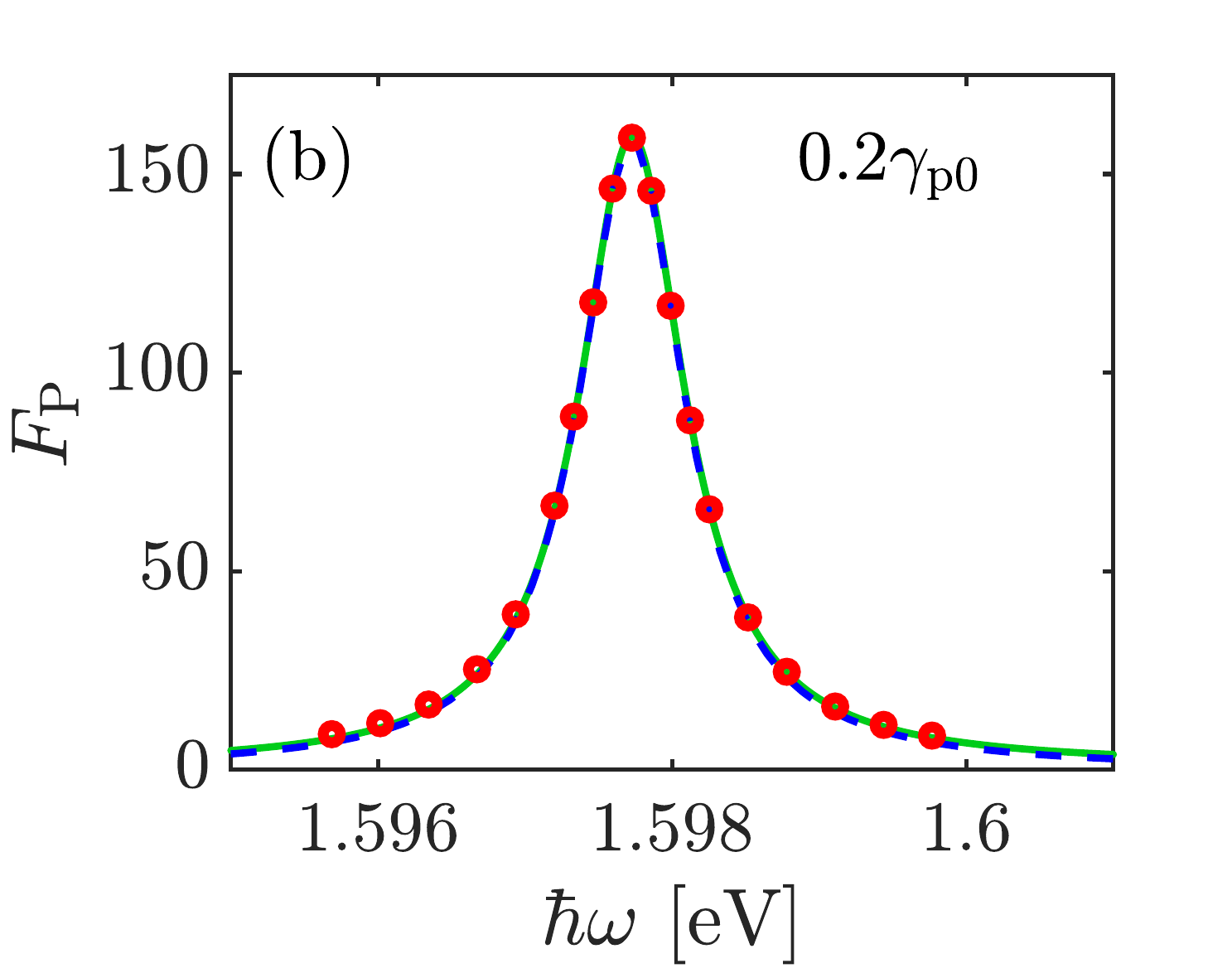}
   \includegraphics[width=0.65\columnwidth]{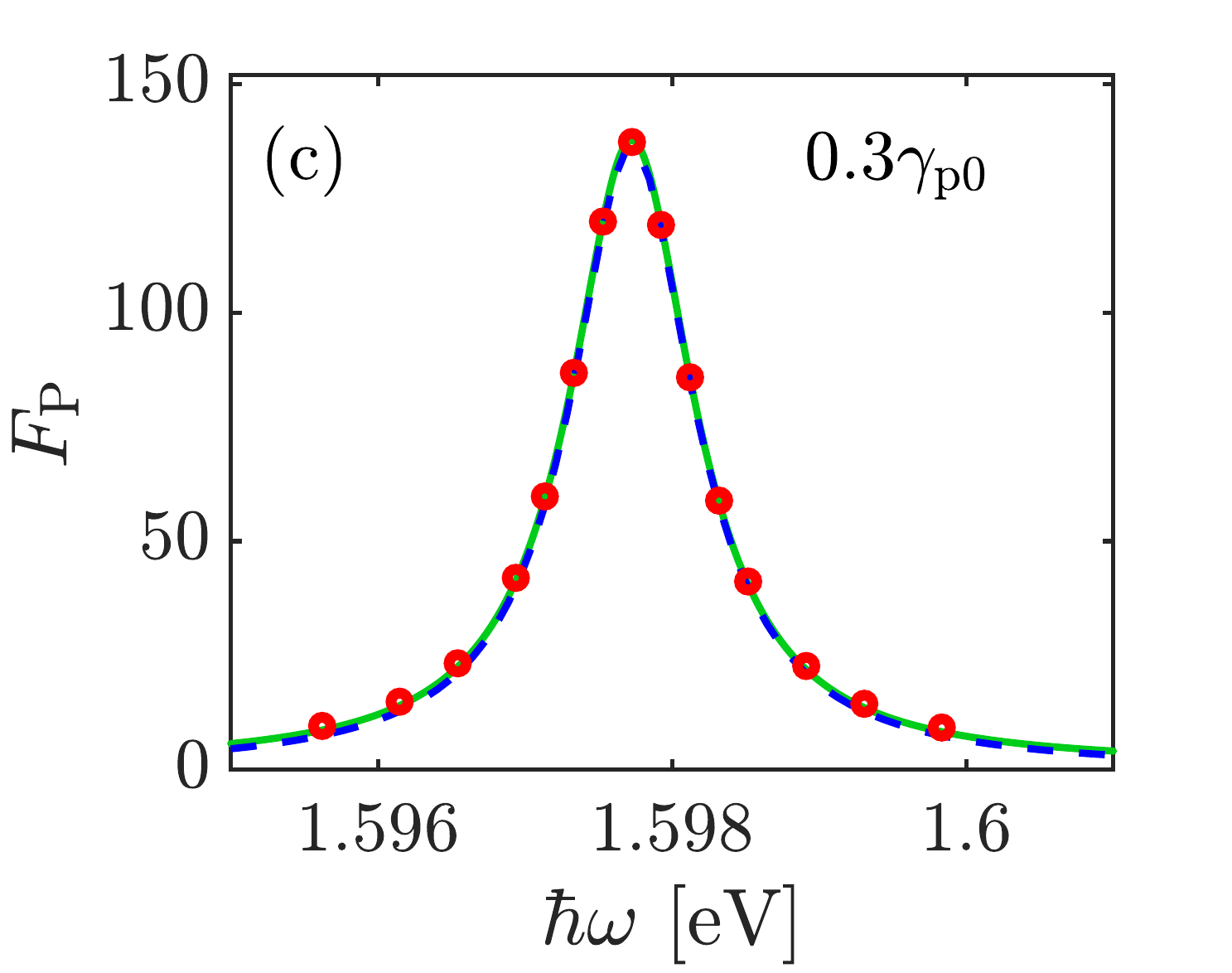}\\
   \includegraphics[width=0.65\columnwidth]{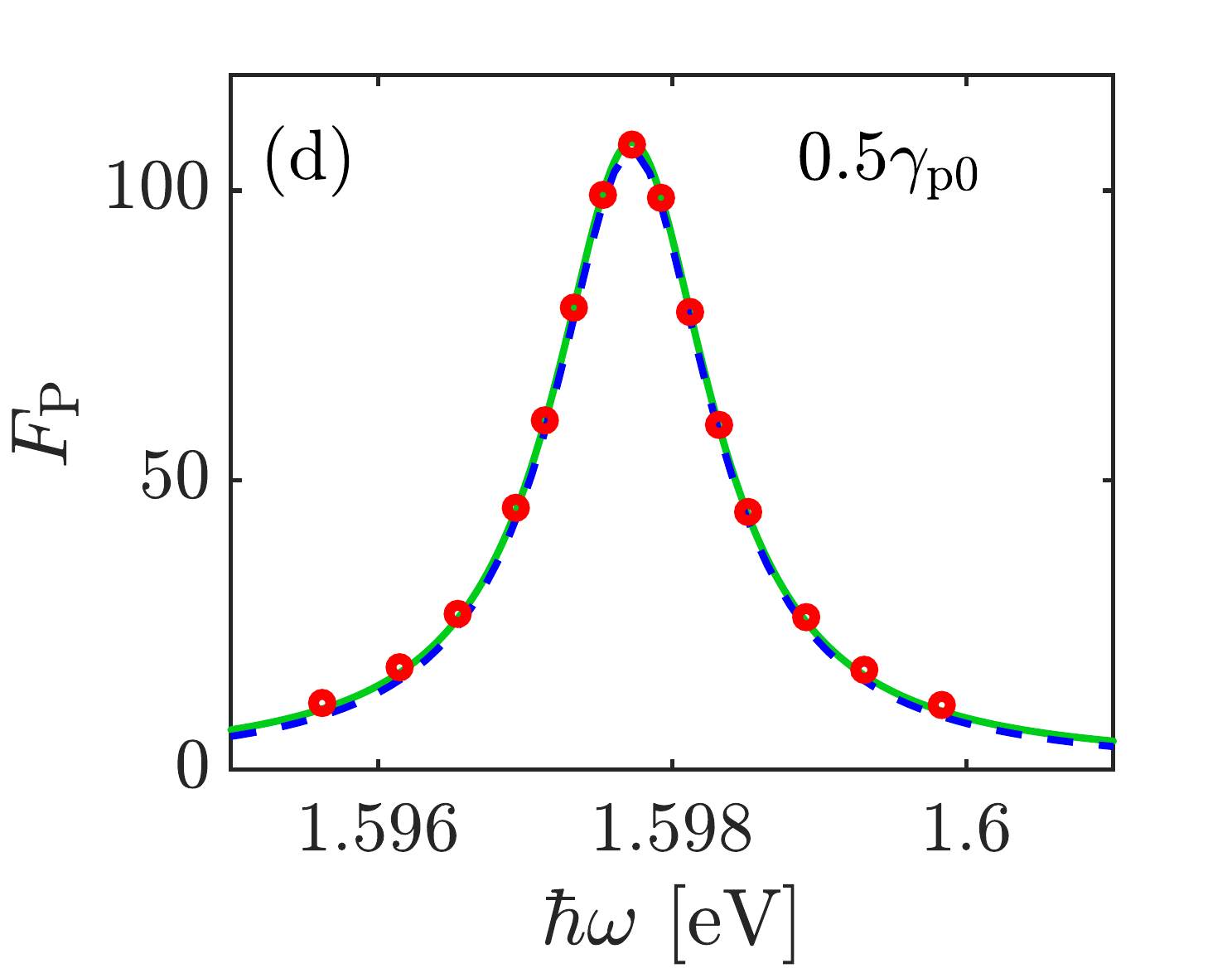}
   \includegraphics[width=0.65\columnwidth]{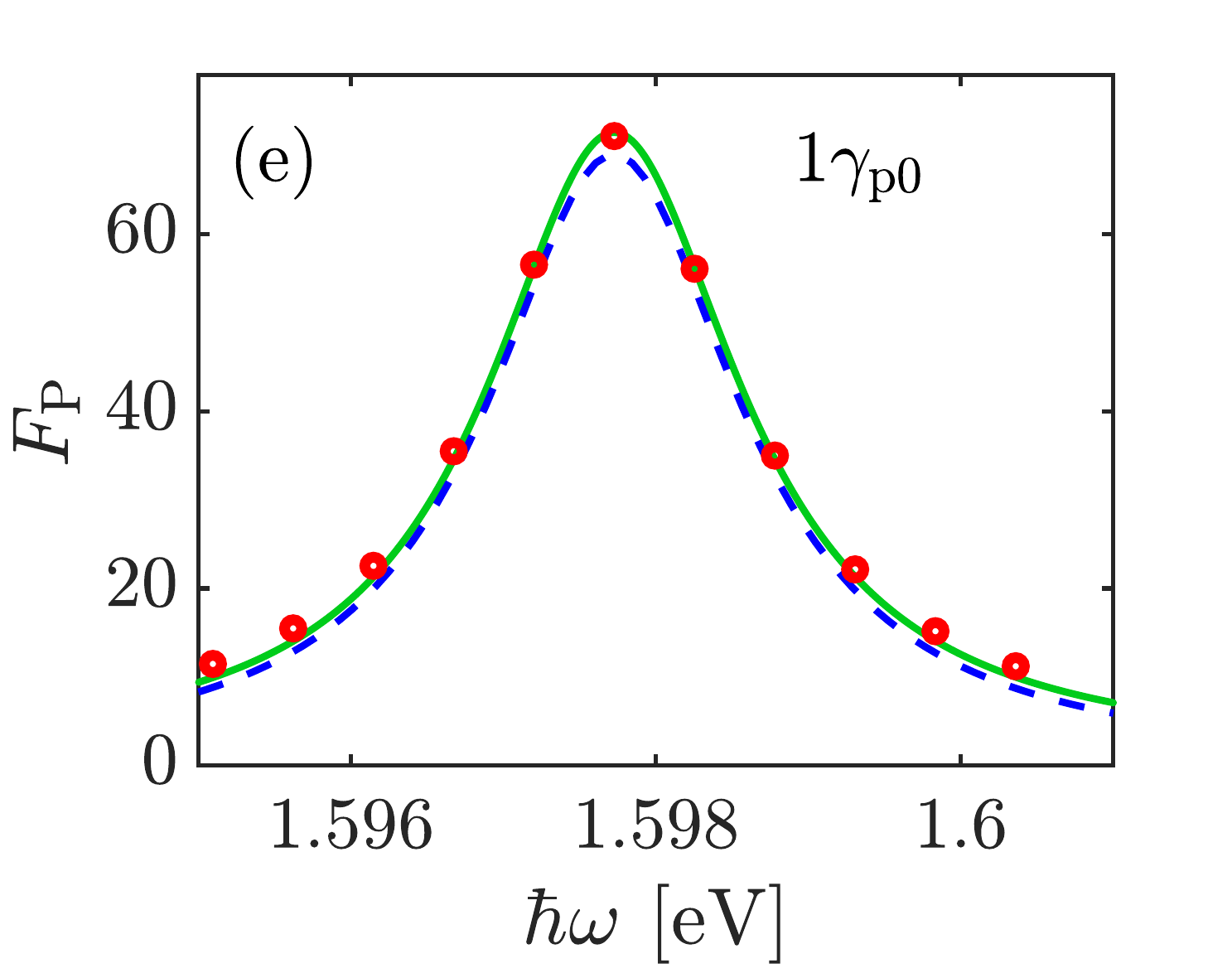}
   \caption{Classical Purcell factor $F_{\rm P}^{\rm QNM}$ (Eq.~\eqref{QNMpurcell}) from single QNM, quantum Purcell factor $F_{\rm P}^{\rm quan}$ (Eq.~\eqref{quantumpurcellsingle}) from quantized QNMs, and numerical Purcell factor $F_{\rm P}^{\rm num}$ (Eq.~\eqref{Purcellfulldipole}) from the full dipole method for a $z$-polarized point dipole placed at $5$ nm above the PC cavity with permitivity of (a) ${\rm Re}[\epsilon_{\rm PC}(1\gamma_{\rm p0})$], (b) $\epsilon_{\rm PC}(0.2\gamma_{\rm p0})$, (c) $\epsilon_{\rm PC}(0.3\gamma_{\rm p0})$, (d) $\epsilon_{\rm PC}(0.5\gamma_{\rm p0})$ and (e) $\epsilon_{\rm PC}(1\gamma_{\rm p0})$. }\label{agree}
\end{figure*}

\begin{figure}[th]
  \centering
  \includegraphics[width=0.9\columnwidth]{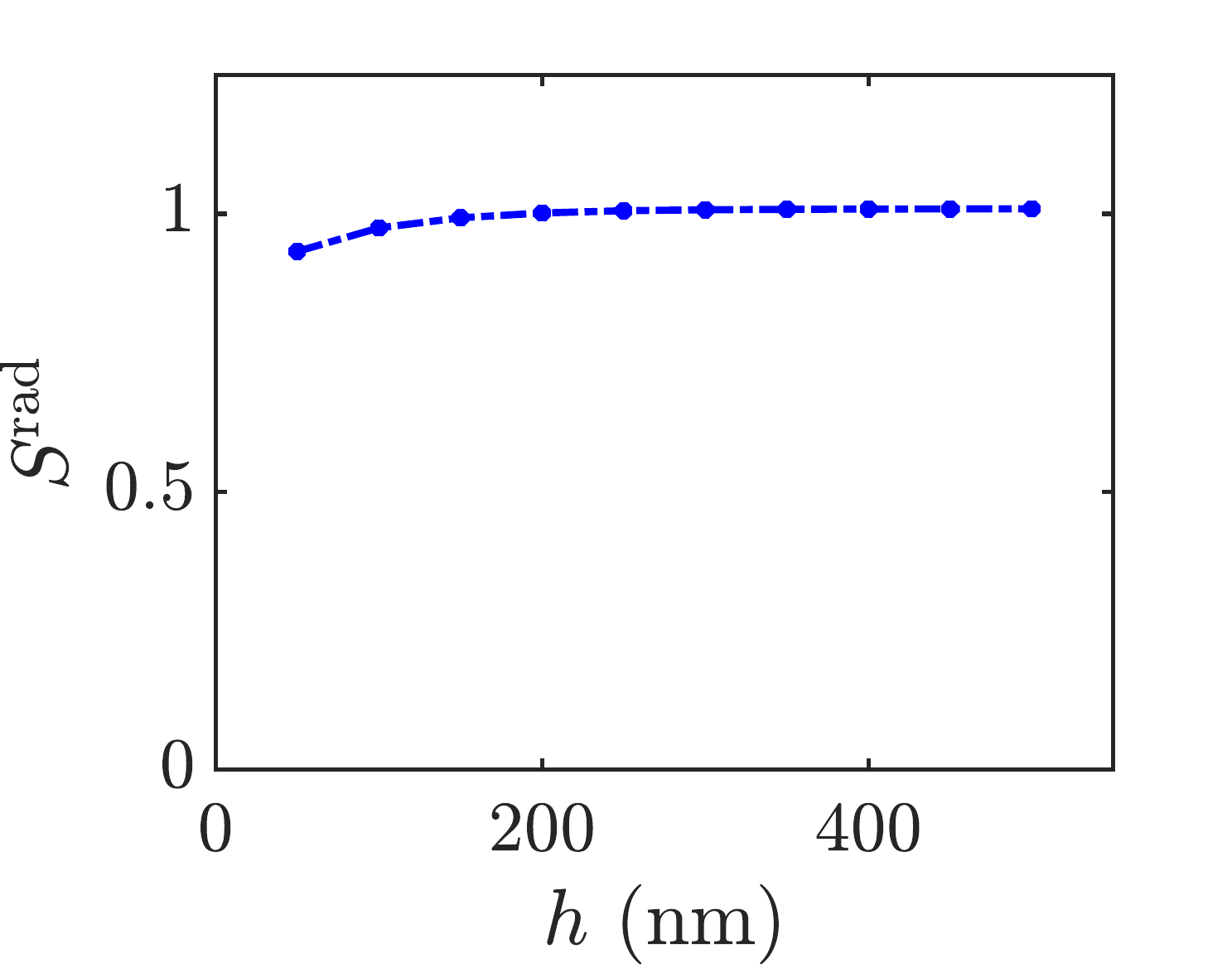}
  \caption{Convergence behavior  of $S^{\rm rad}$ (Eq.~\eqref{srad_pole}) for the lossless PC beam case, as a function of the surface position. These surfaces are specified from cuboid shapes and we use $h_x$, $h_y$, and $h_z$ to represent the distance between these cuboid surfaces and PC beam surface along $x$, $y$ and $z$ direction. We define $h\equiv h_x\equiv h_z$. When $h$ is not larger than $200$ nm, then $h_y=h_x=h_z$. After that, we use fixed value $h_y=200$ nm. Within numerical precision, we found that the radiative $S$ factor for the single QNM tends towards unity, as one might expect for a high $Q$ resonator (see discussions in main text).
  }\label{srad_lossless}
\end{figure}

To confirm the validity of using these regularized fields, we compared the classical Purcell factors $F_{\rm P}^{\rm QNM}$ (Eq.~\eqref{QNMpurcell}) from regularized QNMs with a numerical Purcell factor $F_{\rm P}^{\rm num}$ (Eq.~\eqref{Purcellfulldipole}) from full dipole calculations (i.e., with no approximations). As shown in Figs. \ref{agree} (a)-(e), they fit very well (green curves and red circles).
As highlighted above, when using the normalization method proposed in Ref.~\onlinecite{RegQNMs}, we need to calculate the field at two real frequencies. For all these cases, we choose $\omega_1=\omega_{\rm c}-0.5\gamma_{\rm c}$ and $\omega_2=\omega_{\rm c}+0.5\gamma_{\rm c}$. We also have checked the results using $\omega_1=\omega_{\rm c}-\gamma_{\rm c}$ and $\omega_2=\omega_{\rm c}+\gamma_{\rm c}$, which are nearly the same as those using $\omega_1=\omega_{\rm c}-0.5\gamma_{\rm c}$ and $\omega_2=\omega_{\rm c}+0.5\gamma_{\rm c}$, i.e., this technique is 
robust against the chosen frequency values.

The factor $S^{\rm nrad}$ can be calculated according to Eq.~\eqref{snrad_full} and Eq.~\eqref{snrad_pole} (pole approximation), where a volume integral $I^{\rm spat}_{\rm in}$ over resonator is required.
Note that for all cases with various material losses, the volume integral $I^{\rm spat}_{\rm in}$ are very close to each other (around from $0.1814$ for more lossy case $1\gamma_{\rm p0}$ to $0.1873$ for the completely lossless case). 
Thus, according to Eq.~\eqref{snrad_pole}, the corresponding $S^{\rm nrad}$ will be simply proportional to $Q_{\rm c}{\rm Im}[\epsilon_{\rm PC}(\omega_{\rm c})]$.
We have found the same trend for metallic resonators \cite{ren_near-field_2020},
where further details on the pole approximations are presented.

The factor $S^{\rm rad}$ can be calculated according to Eq.~\eqref{srad_full} and Eq.~\eqref{srad_pole}  (pole approximation), where a surface integral $I^{\rm spat}_{\rm out}$ (surrounding the PC beam) is required.
Here we only focus on the pole approximation. 
For lossless case, we have calculated $S^{\rm rad}$ using regularized QNM at several closed cuboid surfaces (Fig. \ref{srad_lossless}).
We use $h_x$, $h_y$, and $h_z$ to represent the distance between these cuboid surfaces and PC beam surface along $x$, $y$ and $z$ direction.
We define $h\equiv h_x\equiv h_z$.
When $h$ is less than $200$ nm, $h_y=h_x=h_z$.
After that, we use fixed value $h_y=200$ nm.
The grid size is $0.5$ nm for $h=50$ nm, $1$ nm for $100$ nm$\leq h\leq 250$~nm, and $2$ nm for $h\geq300$~nm.
We found $S^{\rm rad}$ converges to a value of $1.009$, close to $1$, when using $h_y=200$ nm, $h_x=h_z$=500 nm with a grid size  of $2$ nm.
So for all lossy cases shown in the main text, we choose the same surface ($h_y=200$ nm, $h_x=h_z$=500 nm) and grid ($2$ nm) to calculate $S^{\rm rad}$.

The factor $S^{\rm rad}$ can also be calculated according to Eq.~\eqref{Sradpole2}, where the near-field to far field transformation is performed and the far field surface is a spherical surface at infinity. The near field surface we choose is the same as above ($h_y=200$ nm, $h_x=h_z=$ 500 nm) and grid size is $2$ nm. We used the same angle grid for integration over both $\vartheta$ and $\varphi$.
We found that $S^{\rm rad\prime}$ from Eq.~\eqref{Sradpole2} converges
when the angle step size is less than
$\pi/35$, and the computed value 
 is very close to those from Eq.~\eqref{srad_pole} (shown in Table~\ref{S1}). This clearly indicates that the regularized field by this approach (normalization method in real frequency space~\cite{RegQNMs}) is indeed 
 practically the same thing {(as $\tilde{\mathbf{F}}$)}, at least at the pole.

Finally, we subsequently obtain $S
 =S^{\rm rad}+S^{\rm nrad}$, and assuming the bad cavity limit, one can calculate the quantum Purcell factors from Eq.~\eqref{quantumpurcellsingle}, which show excellent agreement with results from full dipole method 
 (Figs.~\ref{agree}(a)-(e)).
Note that here the quantum Purcell factors are using $S^{\rm rad}$ obtained from Eq.~\eqref{srad_pole}, and 
we emphasize again, that the results using Eq.~\eqref{Sradpole2} give the same value within the numerical precision.

\end{document}